%% file: TS_QLDPC.tex
\documentclass[a4paper,twocolumn,11pt,accepted=2021-09-28]{quantumarticle}

\pdfoutput=1

\usepackage[utf8]{inputenc}
\usepackage[english]{babel}
\usepackage[T1]{fontenc}

\usepackage[numbers,sort&compress]{natbib}

\usepackage{hyperref}
\usepackage{amsmath,amsfonts,amssymb,wasysym,mathtools}
\usepackage{tabularx,multirow,enumerate,subfigure,bbm,xcolor,url}
\usepackage{wrapfig,sidecap,newproof,eurosym,mathrsfs,algorithmic,algorithm}
\usepackage{stmaryrd,upgreek}
\usepackage{graphicx,epstopdf,epsfig}
\usepackage{bm}
\usepackage{paralist} 
\usepackage{braket}
\usepackage{physics}

\newcommand{\supp}{\mbox{supp}} 

\newtheorem{defn}{Definition}
\newtheorem{te}{Theorem}
\newtheorem{lema}{Lemma}

\newtheorem{ex}{Example}

\newtheorem{remark}[te]{Remark}

\newcommand\sbullet[1][.5]{\mathbin{\vcenter{\hbox{\scalebox{#1}{$\bullet$}}}}}
\newcommand{\sr}{\rule[-0.5cm]{0pt}{1cm}}
\newcommand{\srs}{\rule[-0.4cm]{0pt}{0.8cm}}

\begin{document}
\title{Trapping Sets of Quantum LDPC Codes}
\author{Nithin Raveendran}
\affiliation{Department of Electrical and Computer Engineering, University of Arizona, Tucson, AZ 85721, USA}
\email{nithin@email.arizona.edu}

\author{Bane Vasi\'{c}}
\affiliation{Department of Electrical and Computer Engineering, University of Arizona, Tucson, AZ 85721, USA}
\email{vasic@ece.arizona.edu}
\thanks{\\A part of this paper was presented at the 2021 IEEE International Symposium on Information Theory (ISIT) \cite{ISIT_2021_QuantumTrappingSetNithinConf}.}

\maketitle

	\begin{abstract}
	   Iterative decoders for finite length quantum low-density parity-check (QLDPC) codes are attractive because their hardware complexity scales only linearly with the number of physical qubits. However, they are impacted by short cycles, detrimental graphical configurations known as trapping sets (TSs) present in a code graph as well as symmetric degeneracy of errors. These factors significantly degrade the decoder decoding probability performance and cause so-called error floor. In this paper, we establish a systematic methodology by which one can identify and classify quantum trapping sets (QTSs) according to their topological structure and decoder used. The conventional definition of a TS from classical error correction is generalized to address the syndrome decoding scenario for QLDPC codes. We show that the knowledge of QTSs can be used to design better QLDPC codes and decoders. Frame error rate improvements of two orders of magnitude in the error floor regime are demonstrated for some practical finite-length QLDPC codes without requiring any post-processing. 
    \end{abstract}
   
\input{1Introduction}

\input{2Preliminaries}
\input{3QuantumTS}
\input{4TS_AnalysisCSS}
\input{5_SimulationResultsConclusions}

\section*{Acknowledgment}
We would like to thank David Declercq, Leonid Pryadko, and Saikat Guha for helpful discussions and insights. This work is funded by the NSF under grants CIF-1855879, CIF-2106189, CCF-2100013, CCSS-2052751, and NSF-ERC 1941583.
   

\input{TS_QLDPC.bbl}
\input{6_Appendix}
   
\end{document}

%% file: 1Introduction.tex
\section{Introduction}
\label{Sec:Intro}
Quantum low-density parity check (QLDPC) codes are an important class of quantum error correction (QEC) \cite{mackay_quantum,shor_quantum_memory} codes that can realize scalable fault-tolerant quantum computers (FTQCs) with a finite multiplicative overhead \cite{gottesman_fault_tolerant_ldpc}. In addition, they have finite asymptotic rates with non-zero fault-tolerant thresholds \cite{pryadko_fault_tolerant_ldpc}, and support low-complexity iterative decoding. The existing QLDPC code literature primarily focuses on constructing asymptotically good code families with improved minimum distance scaling with the block length and higher code rates, as well as on designing better iterative decoding algorithms \cite{Hanzo_survey_code, Hanzo_survey_decoders,hypergraphProductCodeTillich,quantum_expander_codes,panteleev2020quantumLinearMinD,refinedBP_QLDPC_2020, kuo2021logdomain}.
However, QLDPC codes implemented in practical QEC systems will be of finite length and will exhibit performance degradation due to the failure of iterative decoders to converge to a correct error pattern. This phenomenon specific to finite-length codes is well understood in classical literature, but similar analysis for QLDPC codes and precise mathematical characterization requires further attention in the QEC literature \cite{Poulin_2008,Hanzo_survey_decoders}. The convergence failure manifests itself as an \emph{error floor} of the decoding probability of error \cite{03Richardson} at low physical error rate levels -- an operating regime for large-scale FTQCs -- and is observed in all state-of-the-art iterative message-passing decoders for QLDPC codes such as belief propagation (BP), min-sum algorithm (MSA) and their variants \cite{Panteleev2019DegenerateQL,VNC_2014_Book,roffe2020decodingQLDPC_OSD}. 

A typical approach in QEC literature to reduce the error floor of the above decoding algorithms is to couple them with \emph{ordered statistics decoding (OSD)} and post-processing \cite{Panteleev2019DegenerateQL,roffe2020decodingQLDPC_OSD}. However, although exhibiting good performance, this technique is too complex to implement in hardware due to the high complexity of the OSD algorithm \cite{ordered_statistic_decoding} which scales cubically with the code dimension 
(see Eq.~4 in \cite{baldi_mrb_decoder_complexity}). In contrast, the philosophy of our approach and our ultimate goal is to develop message-passing decoders for QLDPC codes that do not require a post-processing step to achieve strong error correction.

Iterative message-passing decoder operates on a Tanner graph which is the graphical representation of a parity check matrix of the underlying code.    
Error floor is attributed to the presence of specific topologies of sub-graphs in the Tanner graph, generically referred to as \emph{trapping sets} (TSs) that are detrimental to iterative decoders. Since a trapping set depends both on the topology of the sub-graph and on the decoder, one must understand key differences of QEC, specifically QLDPC codes and decoding with respect to classical error correction. 

The first difference comes as the fact that the stabilizer commutativity/symplectic inner product (SIP) requirement for the parity check matrices introduces additional code construction constraints resulting in unavoidable cycles in the Tanner graph.
Furthermore, QLDPC codes are known to be highly degenerate, i.e., their minimum distance is higher than the weight of their stabilizers. From the decoder perspective, this implies that the decoders need to account for degenerate errors, which have no equivalent in classical error correction. 
However, iterative algorithms based on BP are sub-optimal in the presence of cycles and, also, are not capable of correcting all degenerate errors \cite{Poulin_2008,Rigby2019_ModifiedBP_QLDPC}.  
Another key difference from classical LDPC decoding stems from the inability to directly measure qubits for error correction. Hence, the iterative message-passing algorithms used for decoding of QLDPC codes are modified to only make use of the syndrome information to infer the error introduced by the channel. How the classical trapping sets definition accommodates a syndrome-based decoder is not clearly understood. As we show, degenerate errors having no classical analogy introduces new failure configurations unique to the QLDPC codes. 
The approach presented in this paper accounts for these key differences and their implications.

Failure configurations of QLDPC codes are relatively unknown when compared to the classical trapping set research. One major drawback of BP as pointed out in \cite{Vontobel2019pseudocodewordbased} is that the decoding ability of BP is typically limited by the row weight of the parity-check matrix due to the SIP constraint and identifies pseudo-codeword structures for cycle codes. However, generalization from cycle codes to QLDPC codes is non-trivial.

In this paper, we define quantum trapping sets (QTSs) by investigating failure configurations for syndrome based iterative message passing algorithms. The quantum trapping set formulation is modified to the syndrome decoding scenario for QLDPC codes considering Pauli $\mathrm{X}$ and $\mathrm{Z}$ errors separately. We identify QTSs of prominent QLDPC code families and show that the QTSs must be analyzed in conjunction with the particular iterative decoder used along with their location in the Tanner graph \cite{nithinExp_ContrTCOM2020}. In the same flavor as in classical LDPC codes where the knowledge of trapping sets has resulted in low-complexity decoders surpassing the traditional BP decoder \cite{PDDV_13_COMM}, in this paper, we demonstrate two practical advantages of our quantum trapping set study: (i) ability to construct QLDPC codes devoid of such graphical configurations, and (ii) ability to devise new decoding algorithms that escape from such graphical configurations without post-processing. The message update rules and scheduling strategies thus identified help to improve the error floor performance.

The rest of this paper is organized as follows. In Section \ref{sec:preliminaries}, we introduce QLDPC codes using the stabilizer formalism reviewing some basic notations, and then discuss the syndrome decoding problem and classical trapping sets. In Section \ref{sec:QuantumTS_Procedure}, we analyze the different failure configurations and the relation between trapping sets and decoder-error correction properties. We also formally define quantum trapping sets and describe the methodology used to identify those specifically for Calderbank, Shor, Steane (CSS) codes \cite{calderbank1996quantum_exists}. Trapping sets of some classes of CSS codes are analyzed in Section \ref{sec:TSofCSSCodes}. Based on these analyses, we present simulation results that briefly explore two strategies of code and decoder improvement. We explore CSS code constructions without some of the harmful configurations and compare the performance of trapping set-aware decoding strategies in Section \ref{sec:SimulationResults} followed by concluding remarks and future research directions in Section \ref{sec:Conclusion}.

%% file: 2Preliminaries.tex
\section{Preliminaries}
\label{sec:preliminaries}
\subsection{Stabilizer Formalism}
\emph{Stabilizer} codes, the quantum analog of classical linear codes, are the most common type of QEC codes considered in both theory and practice \cite{Nielsen,Gottesman97}. An $\llbracket n,k,d \rrbracket$ quantum stabilizer code maps $k$ qubit quantum state $\ket{\phi}$ to an entangled $n$-qubit codeword $\ket{\psi}$ (a unit vector in the $2^n$-dimensional Hilbert space
) and is defined as a $2^k$-dimensional subspace of the Hilbert space which is a common $+1$ eigenspace of the stabilizer group $\mathcal{S}$. The $n$-qubit codeword $\ket{\psi}$ is stabilized by all stabilizer elements in $\mathcal{S}$. i.e., $s_j \ket{\psi} = + \ket{\psi}$ for any $s_j \in \mathcal{S}$. We denote a generator set of a given stabilizer group $\mathcal{S}$ by the set $S=\{s_1, s_2,\ldots, s_m\}$. The stabilizer generators form the $m = n-k$ rows of the corresponding stabilizer matrix $H_{p}$ whose entries are the single-qubit Pauli matrices $\mathrm{I_2} = \begin{bsmallmatrix} 1 & 0 \\ 0 & 1 \end{bsmallmatrix}, \mathrm{X} = \begin{bsmallmatrix} 0 & 1 \\ 1 & 0 \end{bsmallmatrix}, \mathrm{Z} = \begin{bsmallmatrix} 1 & 0 \\ 0 & -1 \end{bsmallmatrix}$, and $\mathrm{Y} = \imath \mathrm{X} \mathrm{Z} = \begin{bsmallmatrix} 0 & -\imath \\ \imath & 0 \end{bsmallmatrix}$. Kronecker products of $n$ single-qubit Paulis and scalars $\imath^{\kappa}$, where $\kappa \in \mathbb{Z}_4 = \{0,1,2,3\}$ forms the $n$-qubit Pauli group $\mathcal{P}_n$, of which the stabilizer group $\mathcal{S}$ is a commutative subgroup that contains only Hermitian Paulis and excludes $-I_n$. The \emph{weight} $w(P)$ of a Pauli operator $P \in \mathcal{P}_n$ is the number of qubits on which it applies a non-identity Pauli matrix. A QLDPC code is a stabilizer code with all stabilizer generators having low weight \cite{mackay_quantum}.
Analogous to the classical minimum distance, $\llbracket n,k,d \rrbracket$ code have logical operators $L \in \mathcal{P}_n \setminus \mathcal{S}$ that commutes with all $s_i$ having minimum weight $d$. Like the codeword generators, the logical operators of the code map an $n$-qubit codeword to another. Logical group $\mathcal{L}$ is generated by $k$ logical $\mathrm{X}$ generators: $L_{\mathrm{X}}=\{lx_1, lx_2,\ldots, lx_k\}$ and $k$ logical $\mathrm{Z}$ generators: $L_{\mathrm{Z}}=\{lz_1, lz_2,\ldots, lz_k\}$ obtained by using either Gottesman's \cite{Gottesman97} or Wilde's algorithm \cite{Wilde_logicalOperators_2009}. 

The stabilizers commute with each other following the commutativity relation between two $n$-qubit Pauli operators $P$ and $Q$ defined as follows: $$P \circ Q := \prod_{j = 1}^n P_j \circ Q_j,$$ where $P_j \circ Q_j  = \pm1 \text{ if } P_j Q_j = \pm P_jQ_j.$ The Pauli operators $P$ and $Q$ commute if $P \circ Q = +1$ and anti-commute if $P \circ Q = -1$. Every logical generators commute with the stabilizers, and $Lx_i$ commutes with every other generators except with $Lz_i$ $\forall i \in \{1,k\}$. 

\subsection{Stabilizers as binary parity checks} 
An alternative binary representation maps Pauli matrices to binary tuples as follows: $\mathrm{I_2} \rightarrow (0,0),\; \mathrm{X} \rightarrow (1,0),\;\mathrm{Z} \rightarrow (0,1),\;\mathrm{Y} \rightarrow (1,1)$. More generally, binary representation of an $n$-qubit Pauli operator $P$ will be a binary vector of length $2n$ of the form $\bm{p} = (\bm{p_{\mathrm{X}}}, \bm{p_{\mathrm{Z}}})$, where $\bm{p_{\mathrm{X}}}$ and $\bm{p_{\mathrm{Z}}}$ are of length $n$ each with ones at positions of $\mathrm{X}$- and $\mathrm{Z}$-Pauli components, respectively. Such a mapping aids in the construction of quantum stabilizer codes using extensive classical coding literature. The binary representation $H_{b}$ of the stabilizer matrix of dimension $m \times 2n$ given by
\begin{equation}
H_{b}=\begin{bmatrix}
H_{\mathrm{X}}~|~H_{\mathrm{Z}}
\end{bmatrix},
\end{equation}
where $H_{\mathrm{X}}$ and $H_{\mathrm{Z}}$ represent binary parity check matrices used for error correction. 
Each row in $H_b$ denotes a stabilizer generator, and a pair of corresponding columns in $H_X$ and $H_Z$ represent a qubit. Equivalent to the commutativity relation defined for Pauli operators, the stabilizer generators commute with each other based on the \emph{symplectic inner product} (SIP) in their binary representation \cite{Nielsen}. Any two rows $\bm{p} = (\bm{p_{\mathrm{X}}}, \bm{p_{\mathrm{Z}}})$ and $\bm{q} = (\bm{q_{\mathrm{X}}}, \bm{q_{\mathrm{Z}}})$ of $\begin{bmatrix} H_{\mathrm{X}}~|~H_{\mathrm{Z}} \end{bmatrix}$ must satisfy $\bm{p} \odot \bm{q} := \mod(\bm{p_{\mathrm{X}}}\bm{q_{\mathrm{Z}}}^T + \bm{p_{\mathrm{Z}}}\bm{b_{\mathrm{X}}}^T,2) = 0$. This leads to the condition 
\begin{equation}
\label{SIP_Constraint}
H_{\mathrm{X}}H_{\mathrm{Z}}^T + H_{\mathrm{Z}}H_{\mathrm{X}}^T  = 0,
\end{equation}
where the right hand side ($0$) is an $m\times m$ zero matrix, $T$ denotes the transpose of a matrix, and operations (addition and multiplication) are done modulo-2. We will refer to Eq. \eqref{SIP_Constraint} as the SIP constraint.

\subsection{Decoding Problem}
\label{subsec:DecodingProblem}
Following the approach in \cite{mackay_quantum}, to assess performance of binary syndrome decoding of CSS codes, it is sufficient to consider the two independent binary symmetric channels (BSCs) rather than the depolarizing channel, thus ignoring the correlation between bit flip ($\mathrm{X}$) and phase flip ($\mathrm{Z}$) errors. In this case, the BSCs for $\mathrm{X}$ and $\mathrm{Z}$ errors have a cross-over probability of $2p/3$, decoded using $H_{\mathrm{Z}}$ and $H_{\mathrm{X}}$, respectively.

Let $\bm{e} = (\bm{e_{\mathrm{X}}}, \bm{e_{\mathrm{Z}}})$ be the binary representation of a Pauli error acting on the $n$ qubits. The corresponding syndrome is computed as 
\begin{align*}
\bm{\sigma} &=[ \bm{\sigma_{\mathrm{X}}}, \; \bm{\sigma_{\mathrm{Z}}}]\\ &=[\mod(H_{\mathrm{Z}}.\bm{e_{\mathrm{X}}}^{\text{T}},2),\mod(H_{\mathrm{X}}.\bm{e_{\mathrm{Z}}}^{\text{T}},2) ].
\end{align*}

All-zero syndrome $\bm{\sigma}$ = $\bar{0}$ indicates that all the stabilizers commute with the error pattern (undetectable error), whereas non-zero entries/ones in $\bm{\sigma}$ indicate that some stabilizer generators anti-commute with the error pattern (detectable error). A syndrome based decoder's task is to estimate the error pattern $\bm{\hat{e}}$ whose syndrome $\bm{\hat{\sigma}}$ matches with the initial input syndrome $\bm{\sigma}$. If $\bm{\hat{\sigma}} = \bm{\sigma}$, the estimated error pattern $\bm{\hat{e}}$ is applied to reverse the error $\bm{e}$ introduced by the channel. Error correction process is successful if $\bm{\hat{e}} = \bm{e} \oplus h,$ where $h \in \text{rowspace}(H_b),$ i.e., if the code word is recovered up to a stabilizer ($\bm{\hat{e}} \oplus \bm{e}$ is a stabilizer, where $\oplus$ denotes pairwise XOR). Error correction fails when the decoder is unable to find an error pattern that matches the syndrome $\bm{\sigma}$ or when the decoding process results in a logical error, also referred to as miss-correction in classical coding theory literature. A logical error occurs if $\bm{\hat{e}} \oplus \bm{e}$ is a logical operator such that post error correction state is a codeword different from the original codeword. We can detect a logical error if Pauli representation of $\bm{\hat{e}} \oplus \bm{e}$ anti-commutes with any of the $2k$ logical generators. 

\subsection{Iterative Decoding of CSS codes}
\label{subsec:IterativeDecoding}
Although the syndrome decoding paradigm is applicable to any class of quantum codes, trapping set analysis in this paper is focused on QLDPC families: hypergraph product (HP) codes \cite{hypergraphProductCodeTillich}, bicycle codes \cite{mackay_quantum} and generalized bicycle codes \cite{Panteleev2019DegenerateQL} representing the CSS class of codes \cite{calderbank1996quantum_exists}. An attractive property of CSS codes constructed from two classical codes $\mathcal{C}_1$ and ${\mathcal{C}_2}$, where ${\mathcal{C}}_2^{\perp} \subseteq {\mathcal{C}_1}$ is that the parity check matrix can be written in a separable form: $H_b= \begin{bmatrix}H_{\mathrm{X}}&0\\0&H_{\mathrm{Z}}\end{bmatrix}$. 
CSS-QLDPC codes have a sparse matrix $H_b$ with the SIP constraint: $H_{\mathrm{Z}}. H_{\mathrm{X}} ^\text{T} = 0$. 

We can perform error correction for the ${\mathrm{X}}$ and ${\mathrm{Z}}$ errors separately using $H_{\mathrm{Z}}$ and $H_{\mathrm{X}}$ matrices, respectively. The corresponding input syndromes are obtained as $\bm{\sigma_{\mathrm{X}}} = \mod(H_{\mathrm{Z}}.\bm{e_{\mathrm{X}}}^T,2)$ and $\bm{\sigma_{\mathrm{Z}}} = \mod(H_{\mathrm{X}}.\bm{e_{\mathrm{Z}}}^T,2)$, respectively. For simplicity going forward, we use $H$, $L$ and $\bm{\sigma}$, $\bm{e}$ for the parity check matrix, logical generator matrix, input syndrome, and channel error vector,  respectively.

The stabilizer generator matrix/parity check matrix $H$ is the bi-adjacency matrix of a bipartite Tanner graph $G=(V \cup C, E)$, where $V$ represents the set of $n$ qubit/variable nodes (VNs), $C$ is the set of $m$ stabilizer generators/check nodes (CNs) and $E$ is the set of edges between them. CN $c_i \in C$ and VN $v_j \in V$ are neighbors if there is an edge $(v_j,c_i) \in E$ between the nodes, corresponding to the non-zero entry in the parity check matrix $H_{c_i,v_j} = 1$. Diagrammatically,  
Tanner graphs are drawn with circles representing VNs, squares representing CNs, and solid-lines representing the edges. Let us denote the set of CNs connected to a VN $v_j$ by $\mathcal{N}(v_j)$, and $|\mathcal{N}(v_j)|,$ where $|\cdot|$ denotes cardinality, is referred to as the degree of the VN $v_j$. Similarly, we can define the neighbor set and the degree of a CN $c_i$ as $\mathcal{N}(c_i)$ and $|\mathcal{N}(c_i)|$, respectively. 
A $(\gamma,\rho)$ QLDPC code have a sparse stabilizer matrix with the variable and stabilizer degree upper-bounded by $\gamma$ and $\rho$ respectively. 
For a subset of VNs, say $K \subseteq V,$ $\mathcal{N}(K)$ denotes the set of CN neighbors. The induced sub-graph $\mathcal{G}(K)$ is the graph containing the nodes $K \cup \mathcal{N}(K)$ along with the edges $\{(x,y)\in F : x\in K, y \in \mathcal{N}(K)\}$. 
The girth, $g$, of the Tanner graph $G$ is the length of the shortest cycle in $G$. Denote the number of cycles of length $g$, $g+2, \ldots$ by $\chi_g$, $\chi_{g+2}, \ldots$, respectively.  
If $G$ has $\chi_g$, $\chi_{g+2}, \ldots$ cycles of length $g, g+2, \ldots$, then the cycle enumerator series $\textsc{CYC}(x)=\sum\limits_{r \ge 0} \chi_r x^{r}$ defines the cycle profile of $G$.

The goal of a syndrome-based iterative decoder $\mathcal{D}_s$ is to output an error pattern that matches the input syndrome. This is different from the traditional iterative decoder $\mathcal{D}$ that uses the channel information as initial likelihoods to recover the codeword matching to an all-zero syndrome. Starting from an input syndrome $\bm{\sigma}$ and an all-zero error vector estimate, $\mathcal{D}_s$ performs a finite  number $\ell_{max}$ of iterations of decoding over the Tanner graph. 
The messages are passed over the edges of the Tanner graph from check nodes to their neighboring variable nodes and vice versa at every iteration of message passing decoding. Decoder update rules and message alphabet size can be of varying complexity ranging from the simplest binary message passing algorithms such as Gallager-B \cite{Nithin_stochastic_GallagerB_Quantum} to finite alphabet iterative decoders \cite{PDDV_13_COMM}, and MSA or BP using floating point messages \cite{mackay_quantum}. Also, schedule of message passing in $\mathcal{D}_s$ can be implemented with a  flooding/parallel schedule or a layered/serial schedule. Trapping set analysis presented here is applicable for all such decoder implementations.
We discuss a generic syndrome-based iterative decoder in Appendix \ref{sec:AppendixA} for completeness. 
Based on the update rules, $\mathcal{D}_s$ outputs an error vector estimate $\bm{\hat{e}}^{(\ell)}=(\hat{e}_1^{(\ell)},\hat{e}_2^{(\ell)},\ldots, \hat{e}_n^{(\ell)})$ and corresponding output syndrome $\bm{\hat{\sigma}}^{(\ell)}=(\hat{\sigma}_1^{(\ell)},\hat{\sigma}_2^{(\ell)},\ldots, \hat{\sigma}_m^{(\ell)})$. We refer to $\hat{e}_j^{(\ell)}$/$\hat{\sigma}_i^{(\ell)}$ as the value of the variable/check node $v_j$/$c_i$ at iteration $\ell \le \ell_{max}$. We conclude that $\mathcal{D}_s$ is successful if the output syndrome $\bm{\hat{\sigma}}^{(\ell)}$ is equal to the input syndrome $\bm{\sigma}$ (we also say that syndromes are \emph{matched}). Then, the $n$-length error pattern $\bm{\hat{e}}^{(\ell)}$ is decided as the most likely error pattern. The iterative procedure is halted if successfully matched or if $\ell_{max}$ number of iterations is reached. At the end of iterative decoding, the syndrome decoding process is successful if the syndromes are matched. Otherwise, the decoding is said to have failed.

\subsection{Classical Trapping Sets}
\label{subsec:TSDefinitions}
\begin{figure}[t]
\centering
\includegraphics[width=0.48\textwidth]{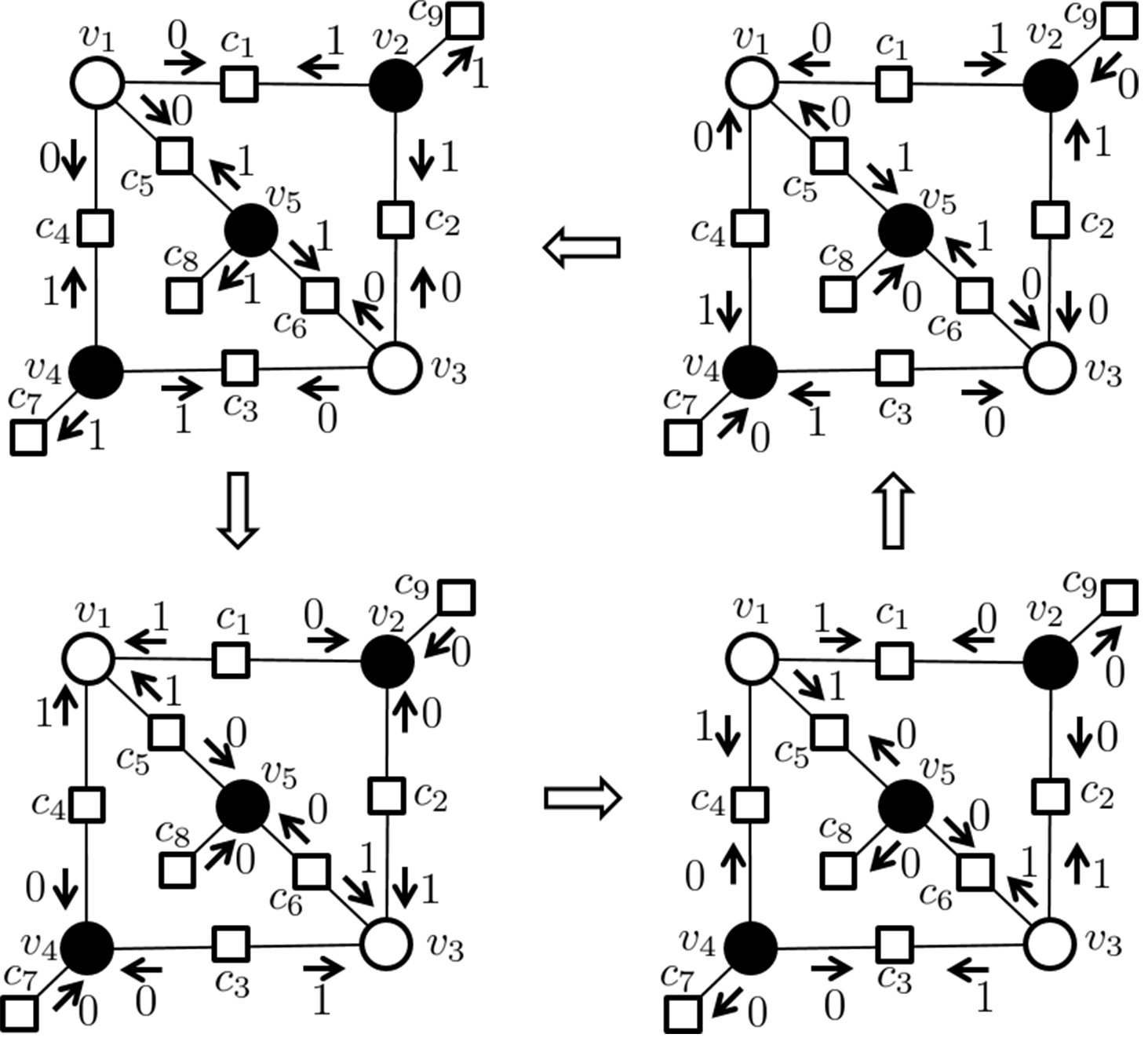}
\caption{An illustration of a failure configuration of regular Gallager-B decoder, unable to converge to the all-zero codeword when the input error pattern is a specific weight-three error pattern ($v_2,v_4,v_5$) (shaded circles $\sbullet[2]$) among the five VNs in the sub-graph. Figures are marked with the binary messages (next to the arrows indicating the direction of the messages passed) corresponding to the check/variable updates. Upper left figure corresponds to the variable node update at the zero-th iteration. The subsequent CN and VN updates of the decoding process are indicated by the connecting arrows. We assume that the rest of the Tanner graph is correct.}
\label{fig_DecodingTrajectory_regular}
\end{figure}

A classical trapping set example is shown in Fig.~\ref{fig_DecodingTrajectory_regular} illustrating the failure of a classical iterative decoder $\mathcal{D}$ on a small sub-graph inside the Tanner graph. Let us consider a simple binary message passing decoder - Gallager-B decoder which performs XOR operation at the check nodes and a majority voting at the variable nodes. More precisely, the outgoing check node message over an edge is computed as the XOR of extrinsic (all incoming messages except the edge for the message is updated) variable node messages. The outgoing variable node message is the majority value among incoming extrinsic check node messages and the channel value. The messages passed over corresponding edges are marked next to the directed arrows in the figure. All-zero transmitted codeword has errors only on three variable nodes ($v_2,v_4,v_5$ - shaded circles ${\sbullet[2]}$) as shown in Fig.~\ref{fig_DecodingTrajectory_regular}. The decoder is unable to converge to the all-zero codeword. In fact, the  decoder oscillates from the error pattern ($v_2,v_4,v_5$) to ($v_1,v_3$) and back as its output during the decoding iterations, thus failing to converge.

A classical iterative decoder is said to converge correctly if the decoder output word for any $\ell \le \ell_{max}$ matches to the transmitted codeword and fails to converge correctly otherwise. A variable node $v_j$ is \textit{eventually correct} if there exists a positive integer $I_j$ such that for all iterations $\ell \geq I_\mathrm{j}$, the decoder's estimate of $v_j$ is equal to the transmitted bit value. Then, trapping set is defined as
\begin{defn}[\cite{VNC_2014_Book}]
\label{Def:TrappingSet}
	A trapping set $\mathcal{T}$ for an iterative decoder $\mathcal{D}$ is a non-empty set of variable nodes in a Tanner graph $G$ that are not eventually correct. If the sub-graph $\mathcal{G(T)}$ induced by such a set of variable nodes has $a$ variable nodes and $b$ odd degree check nodes, then the trapping set $\mathcal{T}$ is conventionally labeled as an $(a,b)$ trapping set.
\end{defn} 

\begin{figure}[t]
\centering
\subfigure[(4,2) TS] 
{
    \label{fig_42}
    \includegraphics[width=0.14\textwidth]{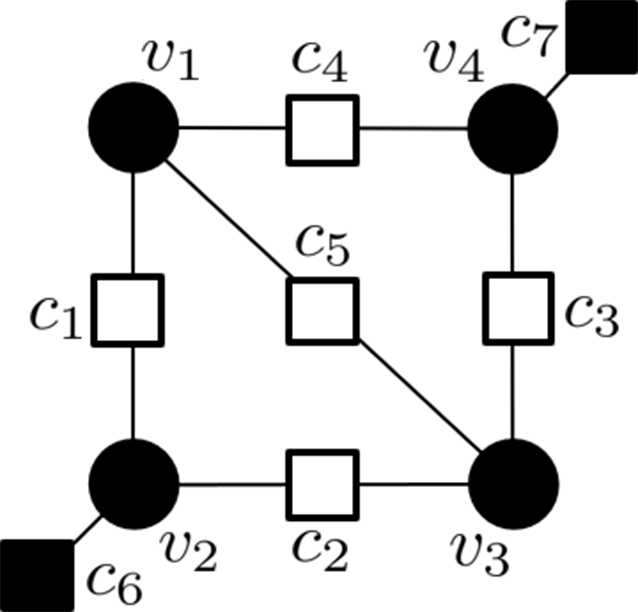}
}
\hspace{0.1in}
\subfigure[(5,3) TS]
{
    \label{fig_53_}
    \includegraphics[width = 0.16\textwidth]{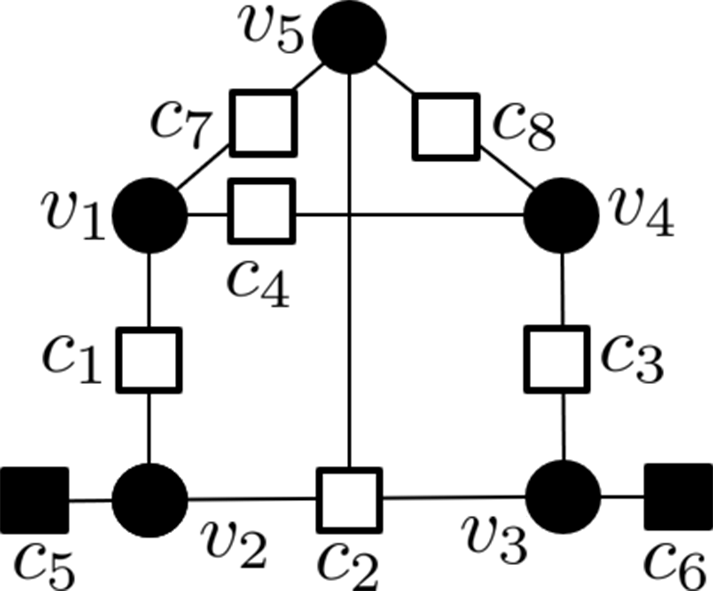}
}
\caption{Graphical (Tanner graph) representations of a $(4,2)$ TS and a $(5,3)$ TS. Odd degree/unsatisfied checks are shown using black squares.
}
\label{fig_Examples_regularTS}
\end{figure}

Fig.~\ref{fig_Examples_regularTS} shows examples of TS induced sub-graphs observed in classical LDPC codes. 
Even at low physical error rate levels, the presence of such small sub-graphs can result in decoding failures resulting in the characteristic error floors in their decoding performance (frame error rate (FER) vs. physical error rate) curves. 

Harmfulness of a TS is also closely linked to the decoder through their critical number $\mu$ and strength $\mathrm{s}$ defined as follows: 
\begin{defn} 
\label{Def:CriticalNumTS}
Critical number $\mu$ of a trapping set $\mathcal{T}$ is the minimal number of variable nodes that have to be initially in error for the decoder to fail to converge.
\end{defn}
Let failure inducing set be the set of variable nodes that have to be initially in error for the decoder to fail to converge. 
\begin{defn} 
\label{Def:StrengthTS}
Strength $\mathrm{s}$ of a trapping set $\mathcal{T}$ is the number of failure inducing sets of cardinality $\mu$.
\end{defn}

Two important assumptions are used in the definition of the critical number and strength of a TS. The first one is that the minimum distance of the code is large compared to the size of the TS. The second is an \emph{isolation assumption} \cite{PDDV_13_COMM} which ensures that messages from outside the TS are correct for the TS failure analysis. 
For example, from Fig.~\ref{fig_DecodingTrajectory_regular}, the critical number $\mu$ for the $(5,3)$ TS with Gallager-B decoder is $3$ and the number of weight-$\mu$ error patterns that fail is $\mathrm{s} = 1$. Note that the decoder also fails to correct weight-4 error patterns and the weight-5 error pattern in the TS. Error floor of the decoder is dominated by the minimum critical number $\mu_{\tiny{\mbox{min}}}$ and the number of weight-$\mu_{\tiny{\mbox{min}}}$ failure inducing error patterns. 
Analytical/semi-analytical estimation of error floors of QLDPC codes using critical number and strength of TSs is beyond the scope of this paper and we refer the reader to classical LDPC literature \cite{VNC_2014_Book,nithinExp_ContrTCOM2020}.

%% file: 3QuantumTS.tex
\section{Quantum Trapping Sets}
\label{sec:QuantumTS_Procedure}
As our focus is on the error floor regime, we are interested in error patterns with small weight, well below the maximum likelihood (ML) error correction capability of the QLDPC code for which the syndrome-based iterative decoder $\mathcal{D}_s$ fails to converge. Such low-weight error patterns are either part of a classical-type TS or a symmetric stabilizer, defined as a quantum trapping set (QTS).

\subsection{Definition of a Quantum Trapping Set}
\label{subsec:TS_definition}
After pre-defined number of iterations, $\ell_{max}$,  of iterative syndrome decoding, we declare that the decoder $\mathcal{D}_s$ failed for a particular input syndrome/error pattern if the decoder is not able to find an error pattern with a syndrome equal to the input syndrome. More precisely, a decoder failure is said to have occurred if there does not exist $\ell \le \ell_{max} $ such that $\supp(\bm{\hat{\sigma}}^{(\ell)}+\bm{\sigma})=\emptyset$, where $\supp$ denotes the support set (indices of non-zero elements).
During iterative decoding, a check node $c_i$ is \textit{eventually satisfied} if there exists a positive integer $I_\mathrm{i}$ such that for all $\ell \geq I_\mathrm{i}$, $\hat{\sigma}^{(\ell)}_i = \sigma_i$. We say that the variable node $v_i$ 
has \emph{eventually converged} if there exists a positive integer $I_\mathrm{i}$ such that for all $\ell \geq I_\mathrm{i}$, $\hat{{e}}^{(\ell)}_i = \hat{{e}}^{(\ell-1)}_i$. Note that the $\hat{{e}}^{(\ell)}_i$ is not necessarily the correct estimate of error on the $i^\text{th}$-variable node. 
With these definitions, we define quantum TSs as follows:
\begin{defn}
\label{Def:TrappingSetSyndromeModified}
	A trapping set $\mathcal{T}_s$ for a syndrome-based iterative decoder $\mathcal{D}_s$ is a non-empty set of variable nodes in a Tanner graph $G$ that are not eventually converged or are neighbors of the check nodes that are not eventually satisfied
\end{defn} 

{\remark{\normalfont If the sub-graph $\mathcal{G(T}_s)$ induced by such a set of variable nodes has $a$ variable nodes and $b$ unsatisfied check nodes, then the trapping set $\mathcal{T}_s$ is conventionally labeled as an $(a,b)$ trapping set.}}

The QTSs similar to the TSs in classical LDPC codes have exactly the same definition as Def. \ref{Def:TrappingSet}, and we refer to them as  \emph{classical-type} trapping sets.
The second class of trapping sets are specifically the harmful degenerate errors observed within the stabilizers classified as \emph{symmetric stabilizer trapping sets}.
We will see that in such trapping sets, even though the variable nodes eventually converge to some error pattern, there exist check nodes that are not eventually satisfied. The definitions and assumptions for critical number and strength of the QTS remain the same as for the classical trapping set.

In the next two paragraphs, we give examples of these two classes of trapping sets. We assume that $\mathcal{D}_s$ is the well-known Gallager-B decoding algorithm. This assumption is made mostly for pedagogical reasons, but also because some trapping sets of Gallager-B are also trapping sets of other decoders such as BP or MSA.

\subsubsection{Classical-type trapping set}
\begin{figure}[t] 
\centering \includegraphics[width=0.48\textwidth]{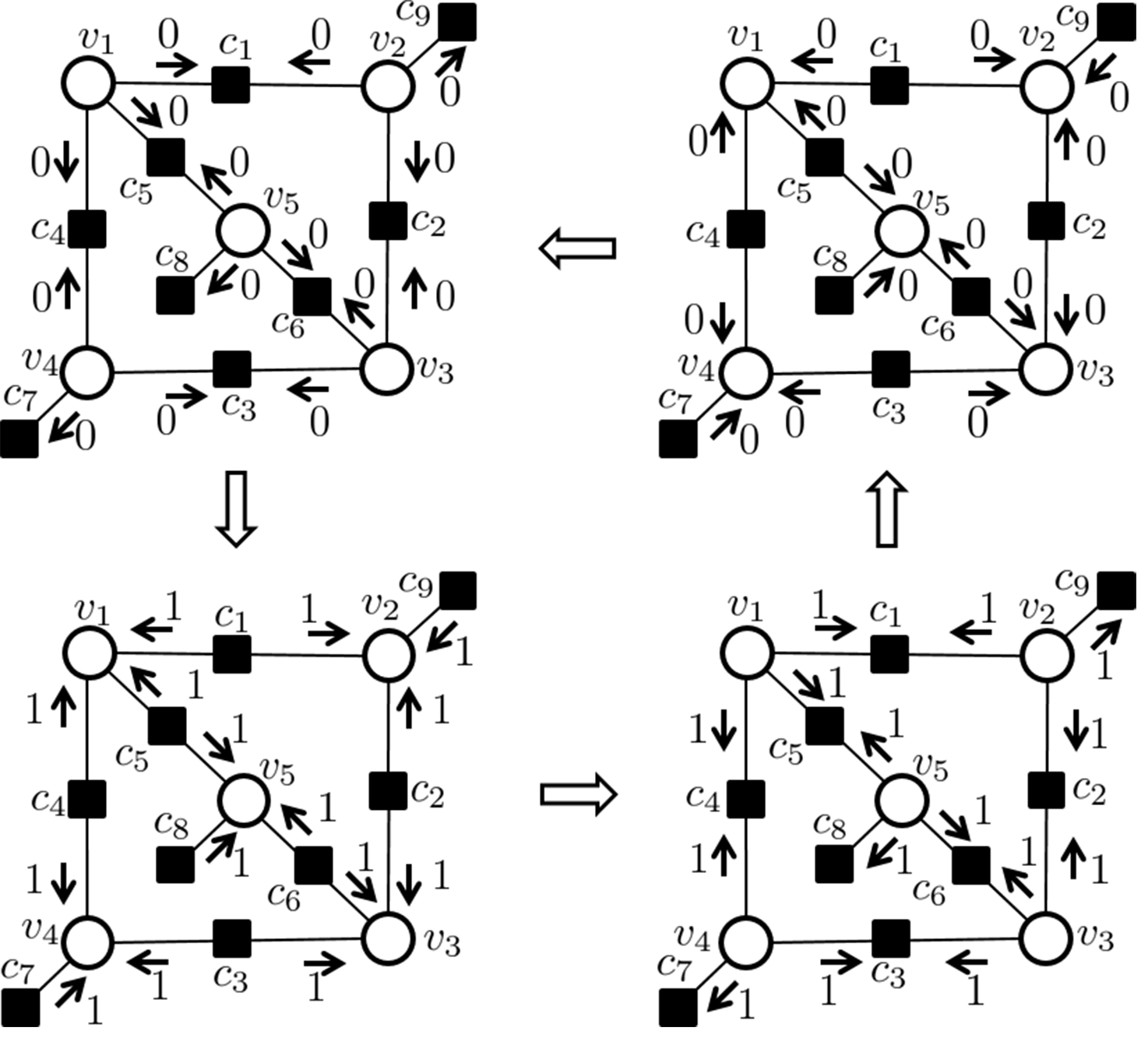} 
\caption{ An illustration of a failure configuration of syndrome-based Gallager-B decoder \cite{Nithin_stochastic_GallagerB_Quantum}. The shaded squares $\blacksquare$ represent the anti-commuting stabilizers/checks. Syndrome iterative decoder starts message passing with an all zero error pattern trying to find the true error pattern that matches with the all-one syndrome but is unable to converge successfully. The messages passed within the sub-graph in consecutive iterations oscillates showing that the decoder is trapped.} \label{fig_DecodingTrajectory_syndrome} 
\end{figure}

We will first show in an illustration why classical-type TSs as shown in Fig.~\ref{fig_Examples_regularTS} are also failure configurations of syndrome decoders. The same error pattern of the (5,3) TS given in Fig.~\ref{fig_DecodingTrajectory_regular} is redrawn for a syndrome based Gallager-B decoder in Fig.~\ref{fig_DecodingTrajectory_syndrome}. The syndrome input is all-one vector, indicated by the black squares and the decoder starts from the all-zero error pattern (no error). The outgoing check node message over an edge is computed as the XOR of extrinsic variable node messages and the syndrome input value at the check node, and the variable node message computation remains the same as in regular Gallager-B decoder. The messages passed over corresponding edges are marked next to the directed arrows. Note that even though the messages passed in Fig.~\ref{fig_DecodingTrajectory_syndrome} are different from that in Fig.~\ref{fig_DecodingTrajectory_regular} of the regular Gallager-B decoder, the syndrome decoder is also unable to converge, and its output oscillates from the all-zero error pattern to errors in $v_2$, $v_4$, and $v_5$ and back. Hence, the $(5,3)$ TS in Fig.~\ref{fig_DecodingTrajectory_syndrome} is classified as a QTS for the syndrome decoder as well.

In addition to classical-type TSs, iterative decoders on QLDPC codes fail for specific degenerate errors. Our quantum trapping set Definition \ref{Def:TrappingSetSyndromeModified} captures such failure configurations as well. This distinctive difference from classical codes deserves further analysis in the next section.

\subsubsection{Symmetric stabilizer trapping set} 
\label{subsec_symmStabilizer}
Recall the quantum decoding problem in Section \ref{subsec:DecodingProblem}, wherein the decoder needs to identify any recovery operator such that $\bm{\hat{e}} \oplus \bm{e} = \text{rowspace}(H_b)$. This is in contrast to the classical decoding problem where an exact match of error $\bm{\hat{e}} = \bm{e}$ is required. In quantum decoding, we say error vectors $\bm{e}$ and $\bm{f}$ are \emph{degenerate errors} if $\bm{e} \oplus \bm{f}$ is a stabilizer, which makes it equivalent to output any one of the degenerate errors as the candidate error pattern for matching the syndrome.
However, in QLDPC codes whose minimum distance is higher than their stabilizer weight, some degenerate errors can be detrimental to iterative decoding. A symmetric topology of the stabilizer sub-graph that contains degenerate error patterns $\bm{e}$ and $\bm{f}$ of equal weight will result in a decoding failure. We will see more examples of such decoder failure when the iterative decoder attempts to converge to error patterns $\bm{e}$ and $\bm{f}$ simultaneously, thus not matching the input syndrome. This failure can be attributed to the symmetry of the both the stabilizer and the decoder message update rules. Hence, such errors are referred to as symmetric degenerate errors and corresponding sets of variable nodes as symmetric stabilizer trapping sets or just symmetric stabilizers, for short. Although degenerate errors are typically classified as harmless for quantum decoding, from the above discussion it follows that some (not all) degenerate error patterns in a symmetric stabilizer are harmful for iterative decoders.
 
\begin{figure}[t]
\centering
\subfigure[] { \label{fig_40} \includegraphics[width = 0.14\textwidth]{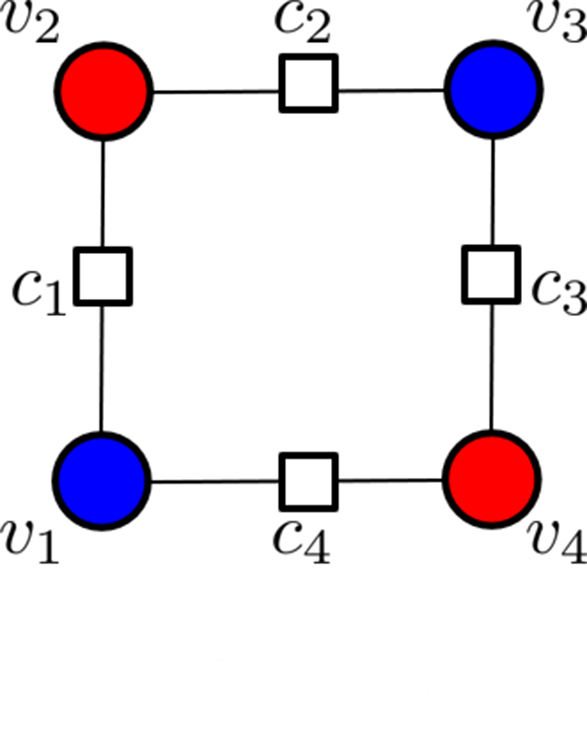} }
\hspace{0.2in}
\subfigure[] { \label{fig_100} \includegraphics[width = 0.22\textwidth]{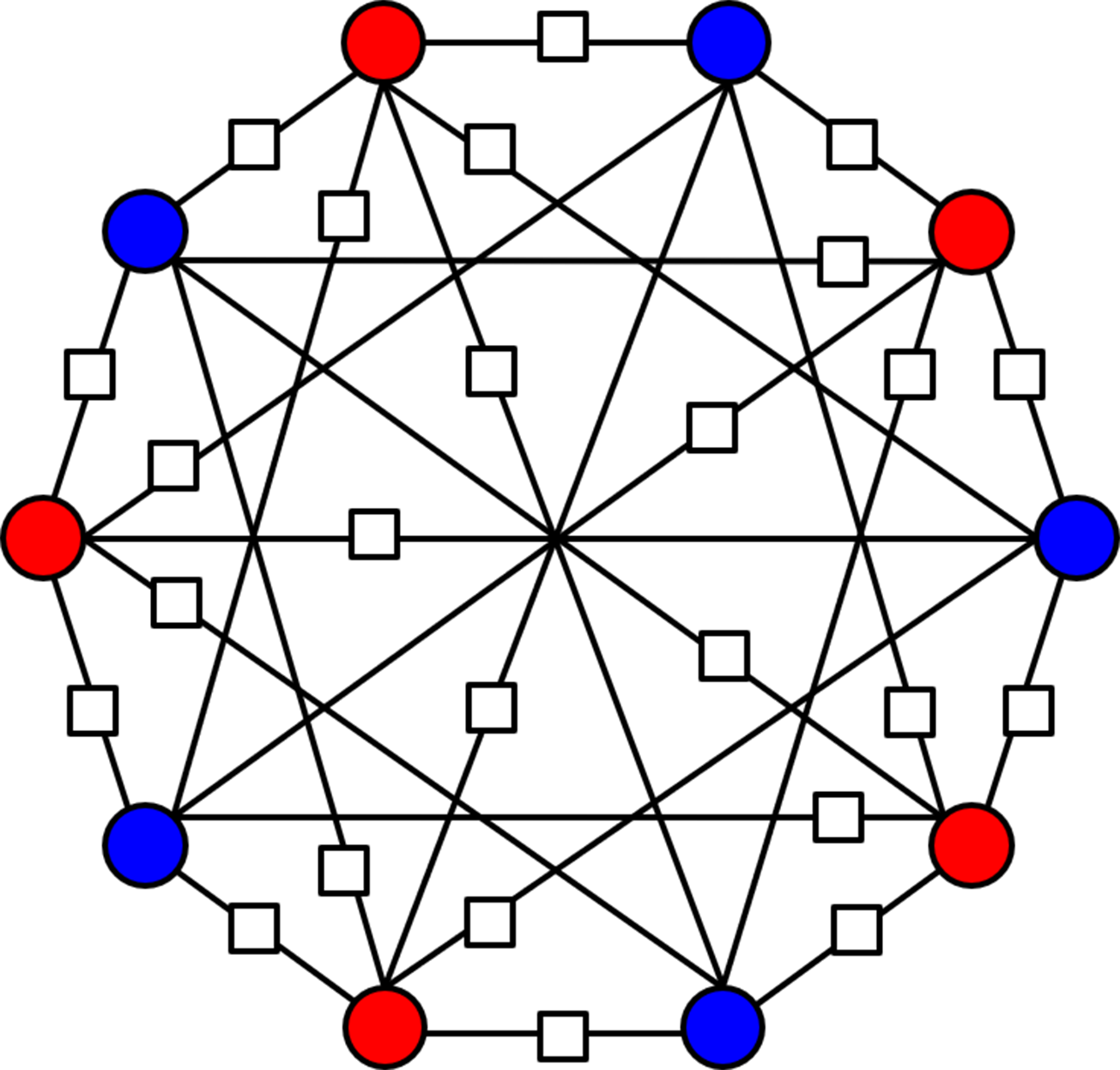} }
\caption{The Tanner graph representations of the $(4,0)$ and $(10,0)$ symmetric stabilizers with $\color{red}{\sbullet[1.5]}$ and $\color{blue}{\sbullet[1.5]}$ representing the disjoint sets of variable nodes of the stabilizer.}
\label{fig_Examples_StabilizersTrivial}
\end{figure}

\begin{defn} \label{Def:SymmStabilizer} A symmetric stabilizer is a stabilizer with the set of variable/qubit nodes, whose induced sub-graph has no odd-degree check nodes, and that can be partitioned into an even number of disjoint subsets, so that: (a) sub-graphs induced by these subsets of variable nodes are isomorphic, and (b) each subset has the same set of odd degree check node neighbors in its induced sub-graph. \end{defn}

\begin{figure}[t]
\centering
\subfigure[] 
{
    \label{fig:red_10_0}
    \includegraphics[width = 0.21\textwidth]{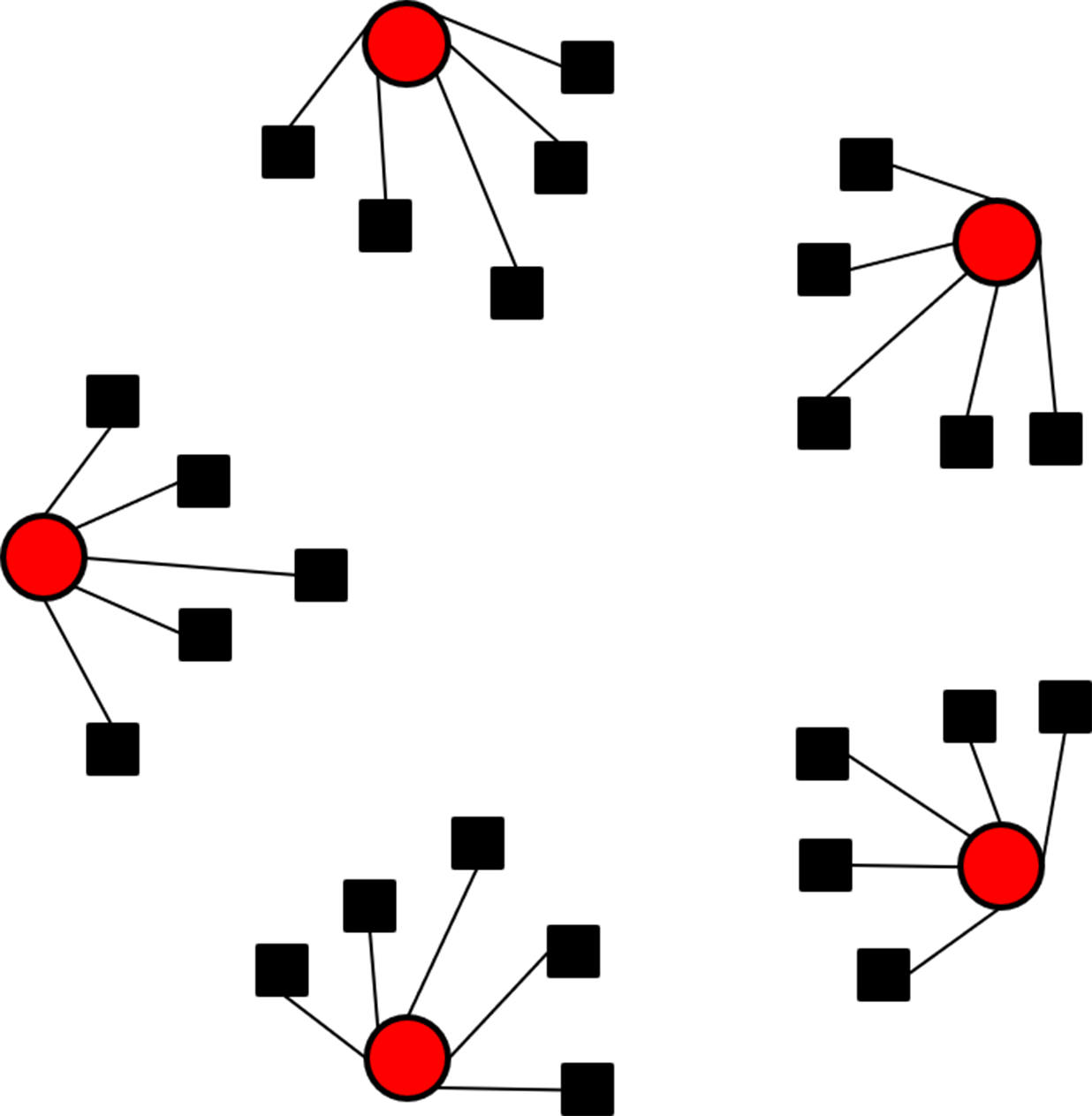}
}
\subfigure[] 
{
    \label{fig:blue_10_0}
    \includegraphics[width = 0.21\textwidth]{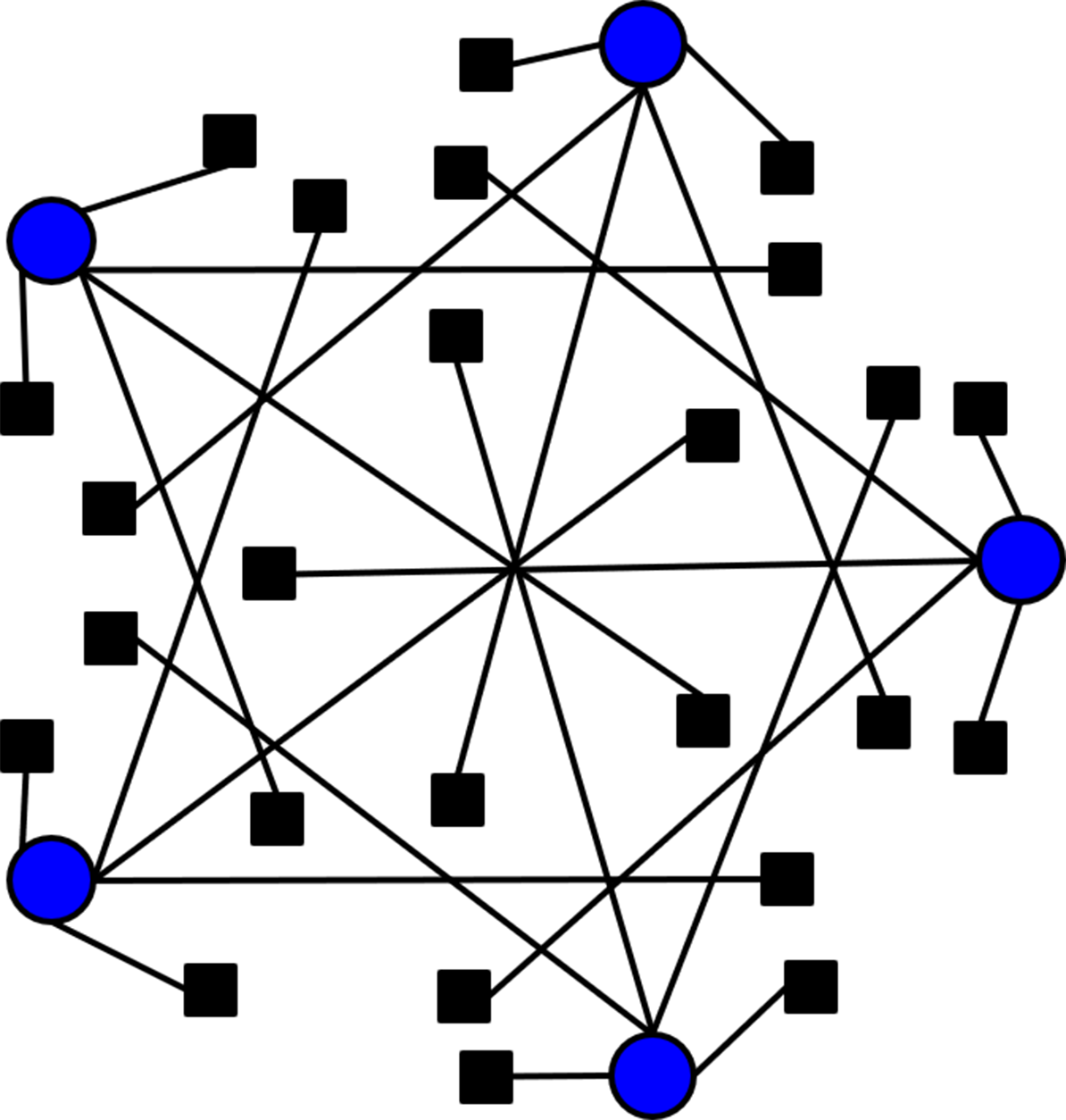}
}

\caption{The induced sub-graphs from $\color{red}{\sbullet[1.5]}$ and $\color{blue}{\sbullet[1.5]}$ variable nodes of a $(10,0)$ symmetric stabilizer trapping set. The sub-graphs 
\ref{fig:red_10_0} and \ref{fig:blue_10_0} are isomorphic and have the same odd-degree checks represented using dark squares $\color{black}{\blacksquare}$.}
\label{fig:SymmStabErrors}
\end{figure}

\begin{ex}
\label{Example:SymmStab}
\normalfont Consider the Fig.~\ref{fig_100} with the stabilizer sub-graph induced by ten variable nodes that are partitioned into two disjoint sets with the coloring $\color{red}{\sbullet[1.5]}$ and $\color{blue}{\sbullet[1.5]}$. The induced sub-graphs from $\color{red}{\sbullet[1.5]}$ and $\color{blue}{\sbullet[1.5]}$ variable nodes are shown in the Fig.~\ref{fig:SymmStabErrors}. The sub-graphs in Fig.~\ref{fig:red_10_0} and Fig.~\ref{fig:blue_10_0} are isomorphic and have the same odd-degree checks represented using dark squares $\color{black}{\blacksquare}$. Hence, the stabilizer shown in Fig.~\ref{fig_100} satisfies the definition of a symmetric stabilizer.
\end{ex}
{\remark{\normalfont The symmetric stabilizer shown in Fig.~\ref{fig_Examples_StabilizersTrivial}\subref{fig_100} is present in generalized bicycle codes given in \cite{Panteleev2019DegenerateQL}. Surface codes are highly degenerate, and symmetric stabilizers, for example as shown in Fig.~\ref{fig_Examples_StabilizersTrivial}\subref{fig_40}, are ubiquitous in them.}}

\begin{figure}[t]
\centering
\subfigure[] 
{
    \label{fig:degenerate_error_a}
    \includegraphics[width = 0.2\textwidth]{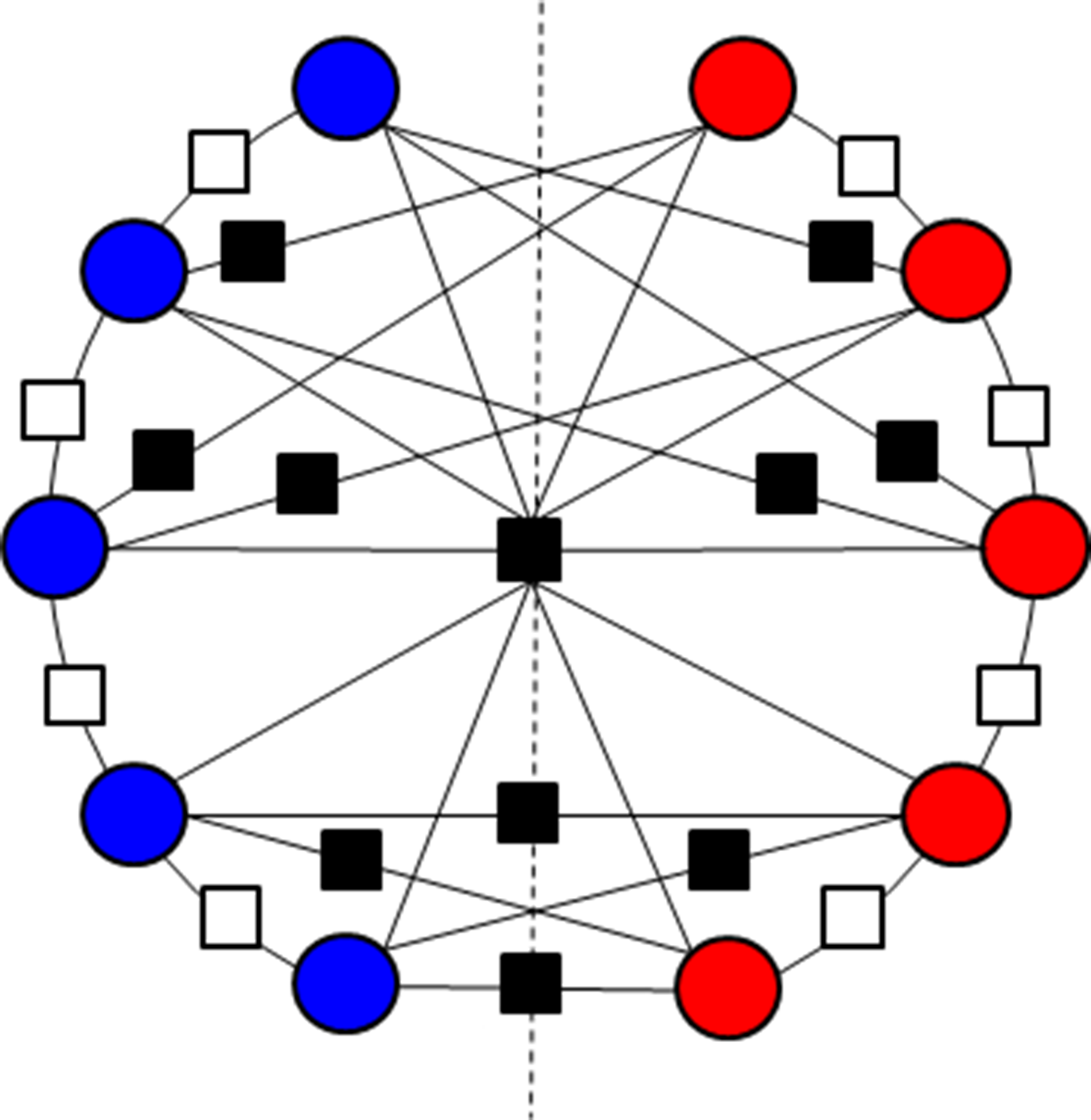}
}
\subfigure[] 
{
    \label{fig:degenerate_error_b}
    \includegraphics[width = 0.2\textwidth]{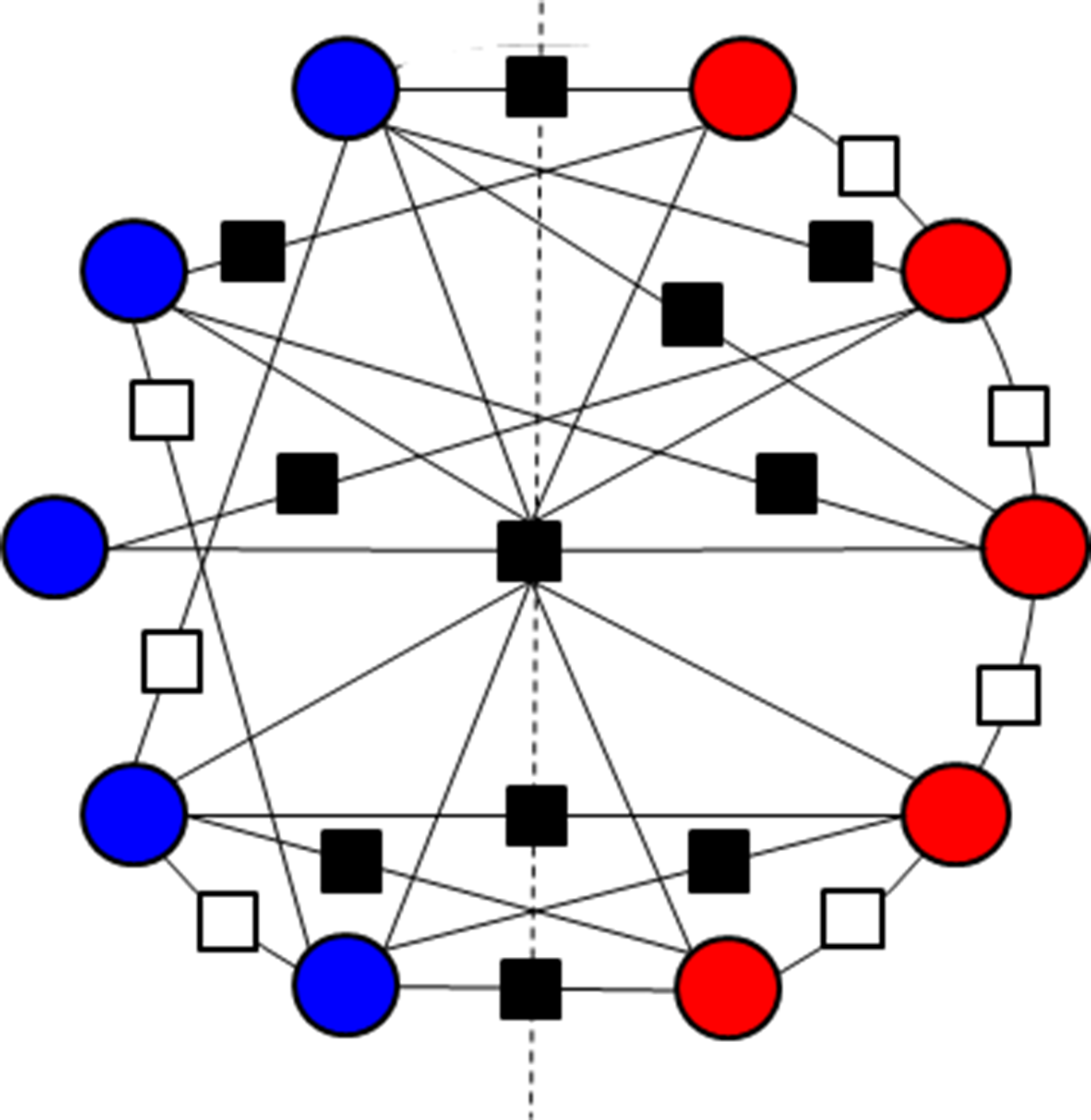}
}
\caption{Degenerate errors $\bm{e}$ and $\bm{f}$ located on $\color{red}{\sbullet[1.5]}$ and  $\color{blue}{\sbullet[1.5]}$ variable nodes, respectively in the symmetric stabilizer in \ref{fig:degenerate_error_a} result in an iterative decoder failure. Introducing asymmetry during the QLDPC code design can lead to decoder success taking advantage of degeneracy of the QLDPC codes. As an example, for the sub-graph in \ref{fig:degenerate_error_b}, a BP decoder is able to match to the syndrome (dark squares represent unsatisfied checks) correctly with the red error pattern.}
\label{fig:degenerate_error}
\end{figure}

Now, we discuss how degenerate errors within the symmetric stabilizer are harmful for iterative decoders. As a non-trivial example of a symmetric degenerate error, let the error pattern $\bm{e}$ be located on the $\color{red}{\sbullet[1.5]}$ variable nodes in Fig.~\ref{fig:degenerate_error}\subref{fig:degenerate_error_a}. They result in unsatisfied check shown as $\color{black}{\blacksquare}$. 
Note, however, that the sub-graph is symmetric with respect to the vertical axis, and therefore each erroneous node has a $\color{blue}{\sbullet[1.5]}$ twin. The set of all $\color{blue}{\sbullet[1.5]}$ twins form an alternative error pattern $\bm{f}$. The existing iterative decoders fail as they simultaneously attempt to converge to both these error patterns. It is not difficult to see that such ``ambiguity'' happens for all decoders for which: (a) the check and message update rules are symmetric functions in incoming messages, and (b) in the same iteration all variable/check nodes in the graph apply in parallel the same variable/check update function, respectively. For example, during the iterations of the Gallager-B decoder, every unsatisfied CN $\color{black}{\blacksquare}$ sends the binary message, one back to the VNs. Because of the symmetry, the VNs in both $\bm{e}$ and $\bm{f}$ receive exactly the same messages, thus converging to $\bm{e} \oplus \bm{f}$, the symmetric stabilizer.  

Based on the Definition \ref{Def:TrappingSetSyndromeModified}, the set of VNs involved in the symmetric stabilizer form a QTS and the sub-graph in Fig.~\ref{fig:degenerate_error}\subref{fig:degenerate_error_a} is a (10,0) TS by convention. We can prove as in the following lemma pertaining to the general case.

{\lema{A symmetric stabilizer is an $(a,b=0)$ trapping set, and $a$ is even. }}
\begin{proof}
Let the cardinality of the set of VNs of the stabilizer be $a$. By definition, the induced sub-graph having no odd-degree check nodes implies that $b = 0$.
Also, according to the symmetric stabilizer definition, the disjoint VN sets that partitions the stabilizer must have the same odd degree check node neighbor set. This implies that there can only be even number of such disjoint sets which further implies that the parameter $a$ is even.
\end{proof}
When there are more than a pair (an even number greater than two) of disjoint sets of VNs, the symmetric stabilizer can be split into smaller symmetric stabilizers.

From the discussion earlier, it follows that symmetric stabilizers are trapping sets not only for the syndrome BP decoder, but for many other iterative decoders, such as bit-flipping, Gallager-B and MSA with different critical number and strength. Harmfulness of symmetric stabilizers associated with decoders is distinct from classical-type trapping sets, as summarized in the following lemma.

{\lema{For an $(a,0)$ symmetric stabilizer TS with any iterative decoder with a critical number $a/2$, no error pattern on more than $a/2$ nodes of the symmetric stabilizer is a trapping set.}}
\begin{proof}
Consider an $(a,0)$ symmetric stabilizer with critical number $a/2$ for a syndrome decoder $\mathcal{D}_s$. By the definition of the critical number, any error pattern of weight smaller than the critical number $a/2$ with support on the symmetric stabilizer is corrected by the decoder $\mathcal{D}_s$. Error patterns of weight larger than the critical number $a/2$ with support on the symmetric stabilizer are decoded correctly, converging to their respective low-weight degenerate error pattern.
\end{proof}

In Fig.~\ref{fig_Examples_StabilizersTrivial}\subref{fig_100}, if a syndrome decoder $\mathcal{D}_s$ is able to correct all error patterns of weight smaller than five, it can also correct error patterns of weight six and more by converging to their respective low-weight degenerate error patterns.

The strength of an $(a,0)$ symmetric stabilizer TS with critical number $a/2$ is given by the twice the number of possible partitions into two disjoint subsets of VNs that satisfy the symmetric stabilizer definition. Each of such partition (distinct by their unsatisfied syndromes) contributes two error patterns each to the decoder failure in the TS. 


\subsection{Searching for Quantum Trapping Sets}
Using the definition of a QTS, one can search for small sub-graphs in the Tanner graph of the QLDPC code to identify and enumerate the QTSs. There are efficient algorithms for TS search widely used in classical literature \cite{Karimi12,DungLatinSquare} to identify sub-graphs that are $(a,b)$ TSs. Such techniques are utilized in the search for classical-type TSs. Note that there can be more than one non-isomorphic sub-graphs with the same $(a,b)$ parameters. For example, a $(5,3)$ TS can have non-isomorphic sub-graphs as in Fig.~\ref{fig_53_} and Fig.~\ref{fig_DecodingTrajectory_syndrome}. Observe that they all have different combinations of short cycles of length six, eight and ten. Enumeration of cycles and their combinations also allows to find harmful classical-type TSs in the QLDPC code.
Unlike these classical-type TSs, the search for symmetric stabilizer TSs requires a different approach of finding low-weight codeword sub-graphs \cite{KDNV_13_ITA} with additional symmetry constraints. In the case of CSS codes, the $H_\mathrm{Z}$ even-weight stabilizer generators are examples of symmetric stabilizer TSs for iterative decoding over the Tanner graph of $H_\mathrm{X}$ matrix and vice-versa. 
After obtaining the list of relevant QTSs, we can perform decoder simulation with an iterative decoder $\mathcal{D}_s$ to verify their relative harmfulness.
In the next section, we find and enumerate QTSs in some prominent QLDPC code families presented in the literature. We also provide the harmfulness analysis of ``dominant'' QTSs present in these code families.

%% file: 4TS_AnalysisCSS.tex
\section{Trapping Set Analysis of CSS codes}
\label{sec:TSofCSSCodes}
A myriad QLDPC code families have been proposed over the years. They include the CSS-based constructions (bicycle codes \cite{mackay_quantum}, hypergraph product (HP) codes \cite{hypergraphProductCodeTillich} and their generalizations \cite{Panteleev2019DegenerateQL}, expander codes \cite{quantum_expander_codes}), non-CSS based QLDPC codes \cite{hanzo_qldpc_from_classical_qc_ldpc,imai_qc_codes} and quaternary QLDPC codes \cite{qldpc_gf_4}. In this section, we analyze trapping sets of CSS based QLDPC codes, the generalized bicycle codes, and HP codes, in particular. Similar analysis may be extended to the general class of stabilizer codes. 

\subsection{Generalized bicycle codes} 
Bicycle codes \cite{mackay_quantum} were generalized by Kovalev and Pryadko in \cite{GeneralizedBicycle_Pryadko} as follows: 
Consider two binary $n/2\times n/2$ matrices $A$ and $B$ that commute ($AB = BA$). Let $$H_{\mathrm{X}} = [A,B] \text{ and } H_{\mathrm{Z}} = [B^T, A^T].$$ The SIP condition is clearly satisfied by definition, and in \cite{GeneralizedBicycle_Pryadko}, $A$ and $B$ are chosen as binary circulant matrices so that they commute. Bicycle codes are dual containing CSS codes where $B = A^T$. Compared to the HP codes, these codes generally have a wider range of parameters; in particular, they can have a higher rate while preserving the estimated error threshold \cite{GeneralizedBicycle_Pryadko}. In \cite{Panteleev2019DegenerateQL}, Panteleev and Kalachev use binary polynomials over rings to define the circulant matrices for constructing $[[n, k]]$ family of generalized bicycle codes. The choice of circulant matrices determines the properties of both the classical-type TSs and the symmetric stabilizers in the code. Hence, the observations on QTSs in the example we discuss next can be generalized to the code family.    
\begin{ex}
\label{ExampleGenBicycle}
\normalfont For illustration, in our TS analysis, we chose the $A1 [[254,28]]$ code, where the circulant size is $127$, $a(x) = 1+x^{15} + x^{20} + x^{28} + x^{66}$ and $b(x) = 1+ x^{58} + x^{59} + x^{100} + x^{121}$ as given in Appendix B in \cite{Panteleev2019DegenerateQL}. The girth of the Tanner graph is six, CN degree $\rho = 10$ and VN degree $\gamma =5$. 
\end{ex}

\subsubsection{Classical-type trapping sets}
\begin{figure}[t]
   	\centering
   	\includegraphics[width=0.47\textwidth]{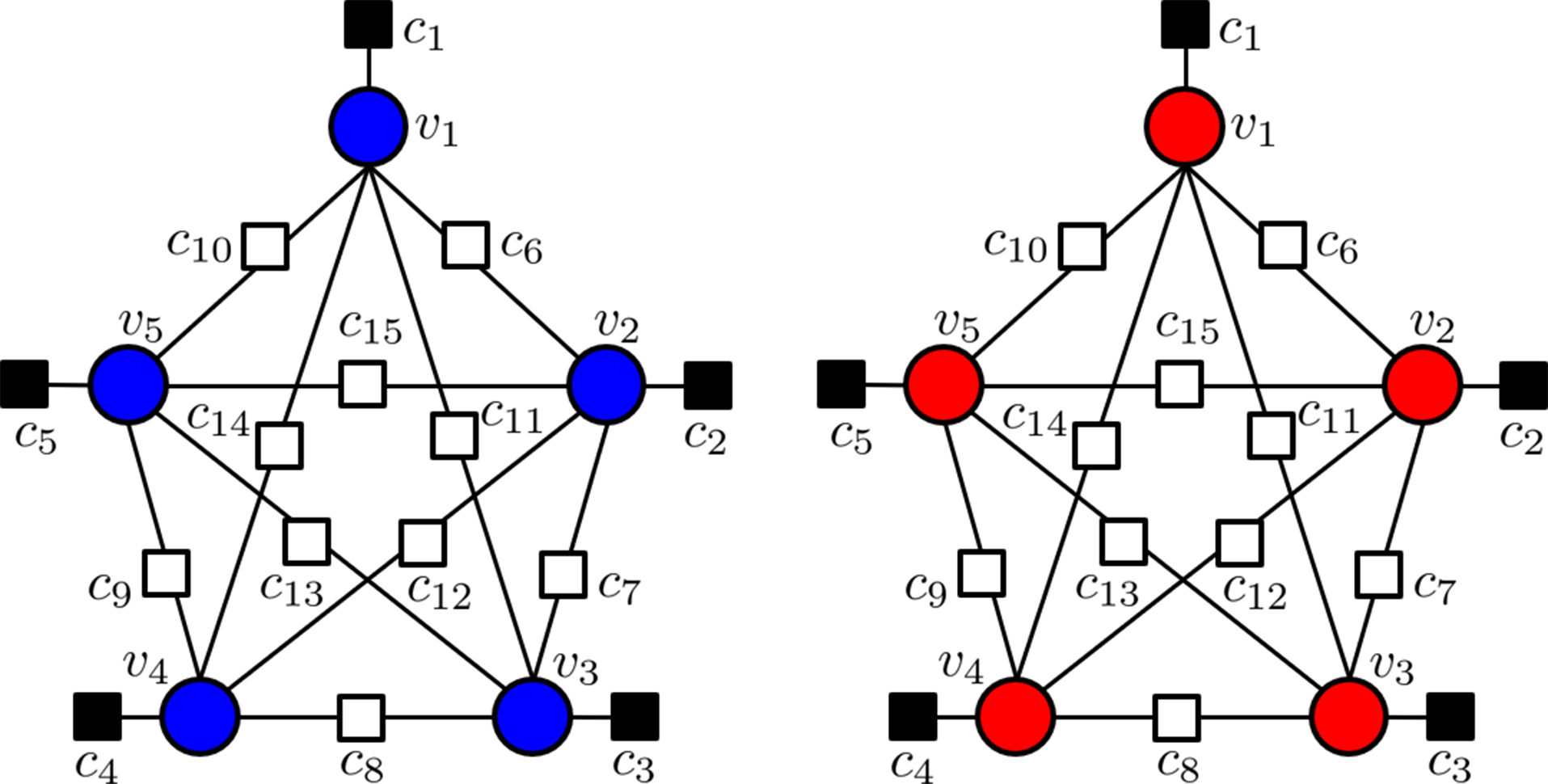}
   	\caption{A $(5,5)$ TS with $5$ variable nodes and $5$ odd degree check nodes (the shaded squares represent the odd-degree checks). The degree of every variable node is $5$. The blue and red shaded circles for the variable nodes indicate their relative position in the $H_{\mathrm{X}}$ matrix, from $A$ and $B$, respectively. }
   	\label{fig:5_5TS_PanteleevsAcode}
   \end{figure}

Based on our QTS definition, we search for  QTSs of small size (upto $a = 5$) present in the $A1$ code in Ex. \ref{ExampleGenBicycle}. As noted in \cite{Panteleev2019DegenerateQL}, based on the circulant matrices in the $A1$ code, it does not have $(a,b)$ trapping sets with $b \le 5$. The $(5,5)$ TS is the most harmful small sub-graph present in the Tanner graph, making BP based iterative decoders to fail for low weight error patterns. The critical number and strength of this TS are determined by the specific decoder used and the neighborhood of the $(5,5)$ TS. Fig.~\ref{fig:5_5TS_PanteleevsAcode} shows the dense $(5,5)$ TS present in both the circulant matrices $A$ and $B$. From their cyclic property, we can locate 127 (equal to the circulant size) isomorphic $(5,5)$ TSs in each of them. In the Fig.~\ref{fig:5_5TS_PanteleevsAcode}, blue and red shaded circles for the VNs indicate their relative position in the $H_{\mathrm{X}}$ matrix, from $A$ and $B$ respectively. 

The $(5,5)$ trapping set in Fig.~\ref{fig:5_5TS_PanteleevsAcode} has five variable nodes. Every variable nodes have exactly the same VN degree $\rho = 5$ and one odd-degree check node neighbor (black squares). The number of small cycles within the trapping set and their symmetry makes this a hard configuration to decode. For simple binary decoders like syndrome based Gallager-B, any weight three or more error patterns will result in a failure inducing set. Hence, the critical number for the $(5,5)$ TS with the Gallager-B algorithm is $\mu = 3$. For stronger decoders such as BP and MSA decoder, the behavior is more complex and interesting. Any weight-$5$ error pattern in the TS in the circulant matrix $A$ indicated by blue qubits results in a failure, whereas similar error patterns in the TSs in the circulant matrix $B$ (indicated by the red qubits) are decoded correctly. Such a behavior is typically attributed to the neighborhood of the TS in the Tanner graph. Whether a TS is  harmful depends not just on the graphical configuration, but also on the neighborhood and the decoder. This ``true'' behavior of all such quantum TSs can be systematically analyzed and characterized by extending the sub-graph expansion-contraction algorithm \cite{nithinExp_ContrTCOM2020} developed for classical LDPC codes to quantum codes, and it is left for future work.

\subsubsection{Symmetric stabilizer trapping sets}  
Failures of syndrome-based iterative decoding on generalized bicycle code also consist of the degenerate error patterns in symmetric stabilizers discussed in the Section \ref{subsec_symmStabilizer}. An example of a pair of symmetric degenerate error patterns of weight five in the Ex. \ref{ExampleGenBicycle} code is shown in Fig.~\ref{fig:red_10_0} and Fig.~\ref{fig:blue_10_0}. Together, they form a $(10,0)$ TS shown in Fig.~\ref{fig_100} referred as a symmetric stabilizer. Interestingly, the blue and red shaded circles indicates the variable nodes relative position as before in case of the $(5,5)$ TS coloring. Also, these error patterns induce isomorphic sub-graphs - trees without any cycles, quite distinct from the error patterns in classical-type TSs which are usually composed of one or more cycles. 
Using the cyclic property of the circulant matrices in the code, we can easily locate $127$ isomorphic symmetric stabilizers present in the Tanner graph of $H_{\mathrm{X}}$, and similarly for $H_{\mathrm{Z}}$. 

{\remark{ \normalfont  The input syndrome (dark squares representing odd-degree checks) in both Fig.~\ref{fig:red_10_0} and Fig.~\ref{fig:blue_10_0} is not matched correctly using iterative decoder using a parallel or flooding schedule. Breaking such TSs requires use asymmetric update of variable node decisions such as used in a layered/serial decoder. Since such symmetric trapping sets have clear distinction of red and blue nodes with respect to the cyclic matrices $A$ and $B$, we can identify the layered decoder schedule that can break such trapping sets.}}

\subsection{Hypergraph product codes}
HP codes by Tillich and Zemor  \cite{hypergraphProductCodeTillich} and their improvements by Kovalev and Pryadko \cite{hypergraphProductCodekovalev_pryadko_improved} are constructed by taking Kronecker product (denoted as $\otimes$) of two classical LDPC codes. 
Using two classical parity check matrices $H_1$ and $H_2$ of dimensions $m_1 \times n_1$ and $m_2 \times n_2$ respectively, we have  
$H_{\mathrm{X}} = \begin{bmatrix} H_1 \otimes \mathrm{I_{n_2}} \; | \; \mathrm{I_{m_1}} \otimes H_2^T \end{bmatrix}$ and $H_{\mathrm{Z}} = \begin{bmatrix}  \mathrm{I_{n_1}} \otimes H_2 \; | \; H_1^T \otimes \mathrm{I_{m_2}}  \end{bmatrix}.$ 

\begin{ex}
\label{ExampleHGP}
\normalfont For our trapping set analysis, we use the example of a $[[900,36,10]]$ HP code given in \cite{roffe2020decodingQLDPC_OSD} using a symmetric Kronecker product of a single $(n= 24,k=6,d=10)$ classical code.  
\end{ex}
The classical LDPC code determines the HP code properties and its trapping sets. As we analyze the cycle profile of the Tanner graph of the classical code used in Ex. \ref{ExampleHGP}, we observe that there are $54$ cycles of length six ($6$-cycles) and $160$ $8$-cycles. 

\begin{table}[h]
	\caption{HPG code parameters and number of cycles}
	\label{table:CodeParams}
	\centering
	\begin{tabular}{|c|c|c|c|c|c|c|}
	\hline 
		\textbf{Code} & $n$ & $m$ &
		\textbf{$g$} & $\chi_g$   &$\chi_{g+2}$\\
		\hline
		[24,6,10] & 24 & 18 & 6 & 54 & 160 \\
		\hline
		[[900,36,10]] & 900 & 432 & 6 & 2268 & 14496 \\
		\hline
	\end{tabular}
\end{table}

These cycles appear in the HP codes, multiplying according to the size of the classical parity check matrix ($m = 18, n = 24$) as given in Table \ref{table:CodeParams}. For example, $54$ six cycles in the classical code gives rise to  $(54 \times 24) + (54 \times 18) = 2268$ $6$-cycles in both $H_{\mathrm{X}}$ and $H_{\mathrm{Z}}$ matrix of the $[[900, 36,10]]$ code. This behavior is consistent across the symmetric HP code family.
{\lema{Tanner graphs of hypergraph product codes have girth at most 8.}}
\begin{proof}
The Kronecker product of the constituent LDPC code graphs results in unavoidable 8-cycles in the Tanner graphs of $H_{\mathrm{X}}$ and $H_{\mathrm{Z}}$ matrices, upper bounding the girth of HP codes by 8 \cite{hypergraphProductCodeTillich}. Let the two constituent code Tanner graphs be $(V \cup C)$ and $(V’ \cup C')$, then one can find a cycle of length $8$ in the Tanner graph of $H_{\mathrm{X}}$ by looking at $H_{\mathrm{Z}}$ rows. Consider a variable node $v' \in V'$ and a check node $c \in C$. Let $v’$ be connected to check nodes $c_1'$  and $c_2'$ in the graph $(V’\cup C’)$. Similarly, let the check node $c$ has variable node neighbors $v_1$ and $v_2$ in the graph $(V \cup C)$. Based on the Kronecker product, the following length-8 cycle is formed involving variable node $(v_1, v’)$ in the hypergraph product: 
$(v_1, v’)$-$(v_1, c_1')$-$(c,c_1')$-$(v_2,c_1')$-$(v_2,v')$-$(v_2,c_2')$-$(c,c_2')$-$(v_1,c_2')$-$(v_1,v')$.
\end{proof}

Our TS search procedure identified thirty $(4,2)$ trapping sets and ten $(5,1)$ trapping sets in the classical code. Also, there are two non-isomorphic topologies of $(5,3)$ TSs: one hundred and seventy $(5,3)$ TSs whose induced graph has a six, eight, and ten-cycle, and fifteen $(5,3)$ TSs having three eight cycles. All these trapping sets manifest themselves in the HP code with their count scaling as in the case of small cycles. In Table \ref{table:TSs}, we enumerate all the smallest QTSs (with $a\le5,b\le a$) present in the $[[900,36,10]]$ HP code having no CN with degree $>2$ in their induced sub-graphs. These values for $a$ and $b$ are chosen as such classical-type TSs are typically the most harmful for iterative decoders. The QTS enumeration of $H_{\mathrm{X}}$ and $H_{\mathrm{Z}}$ for symmetric HP codes is the same. Also, the cycle enumerator series $\textsc{CYC}(x)=\sum\limits_{r \ge 0} \chi_r x^{r}$ for each QTS sub-graph in the parameters column in the Table indicates the number of small cycles present, which we refer to as its cycle profile. Observe that the $(5,3)$ QTS with $\gamma=3$ has two non-isomorphic topologies with different cycle profiles. Another interesting observation specific to these HP codes having VNs with degree $3$ and $4$ is the presence of $(5,3)$ and $(5,5)$ QTSs with the same cycle profile - $\textsc{CYC}(x)=3x^8$. The $(5,5)$ QTSs are indeed the result of the Kronecker product in the HP codes. The QTSs in Table \ref{table:TSs} are the main reason for poor iterative decoding performance of such family of codes.\\

\begin{table*}[t] 
\caption{QTS enumeration in $H_{\mathrm{X}}$/$H_{\mathrm{Z}}$ of $[[900,36,10]]$ HP code  \cite{roffe2020decodingQLDPC_OSD}} 
  \label{table:TSs}
  \centering
\begin{tabular}{|c|c||c|c|}
\hline 
\multirow{4}{*}{\textbf{Quantum TS}} & \textbf{Parameters} \footnotemark & \multirow{4}{*}{\textbf{Quantum TS}} & \textbf{Parameters} \\ \cline{2-2} \cline{4-4} 
& (a,b) && (a,b) \\ &$\textsc{CYC}(x)$ &&$\textsc{CYC}(x)$ \\ & Count &&Count\\ 
\hline 
\multirow{3}{*}{\centering \includegraphics[width=6em]{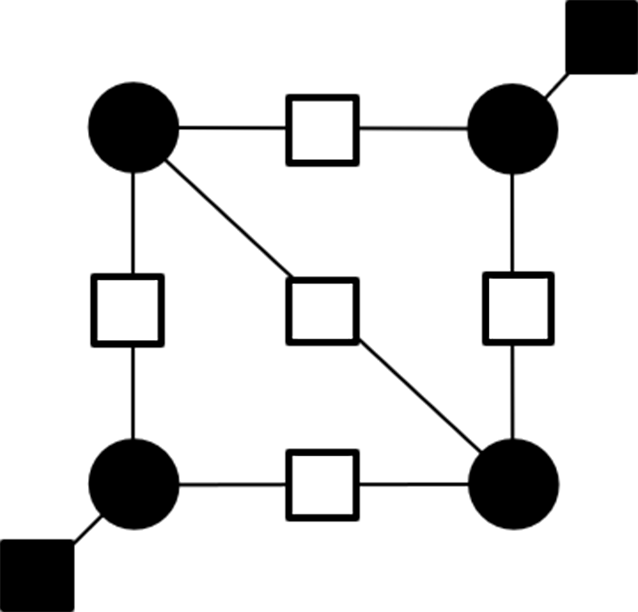}} &\srs (4,2)  &\multirow{3}{*}{\centering \includegraphics[width=6em]{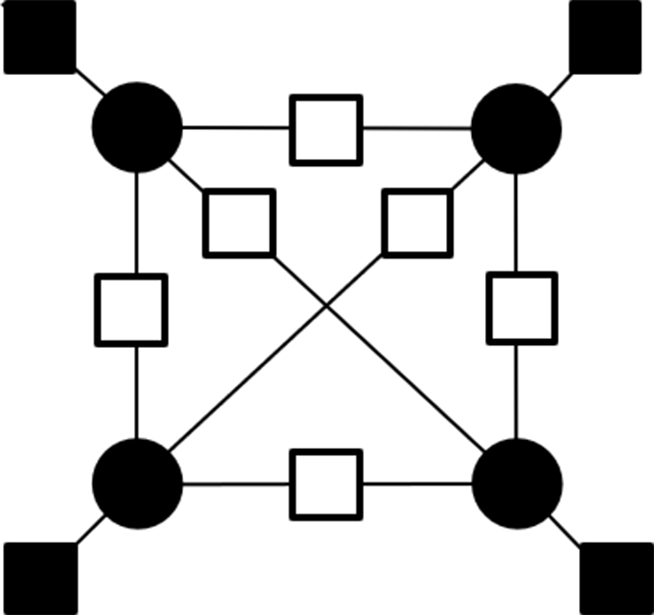}}  &\srs (4,4) 
\tabularnewline
&\srs $2x^6+x^8$  &&\srs $4x^6+3x^8$ 
\tabularnewline
&\srs 720 &&\srs 72 
\tabularnewline
\hline 
\multirow{3}{*}{\centering \includegraphics[width=5em]{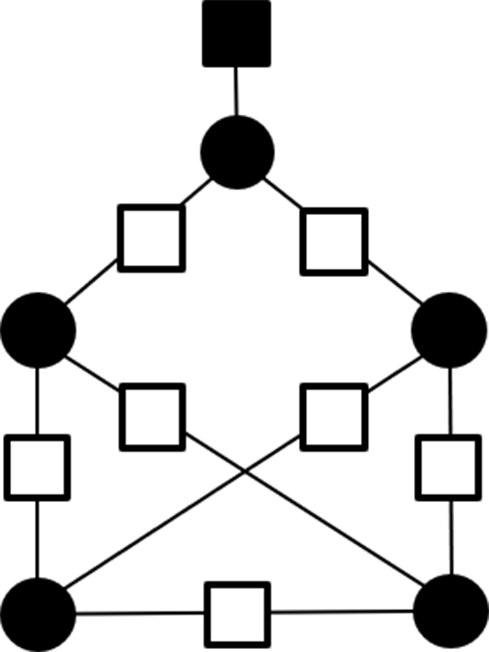}} &\sr (5,1) &\multirow{3}{*}{ \centering \includegraphics[width=7.5em]{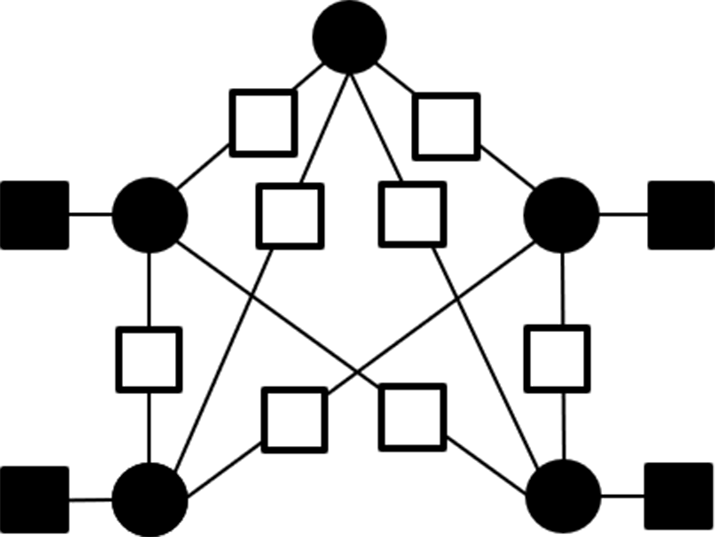}} &\sr (5,4) 
\tabularnewline
&\sr $2x^6+3x^8+2x^{10}$ && \sr $4x^6+5x^8+4x^{10}$ 
\tabularnewline
&\sr 240  &&\sr 36
\tabularnewline
\hline 
\multirow{3}{*}{\centering \includegraphics[width=7em]{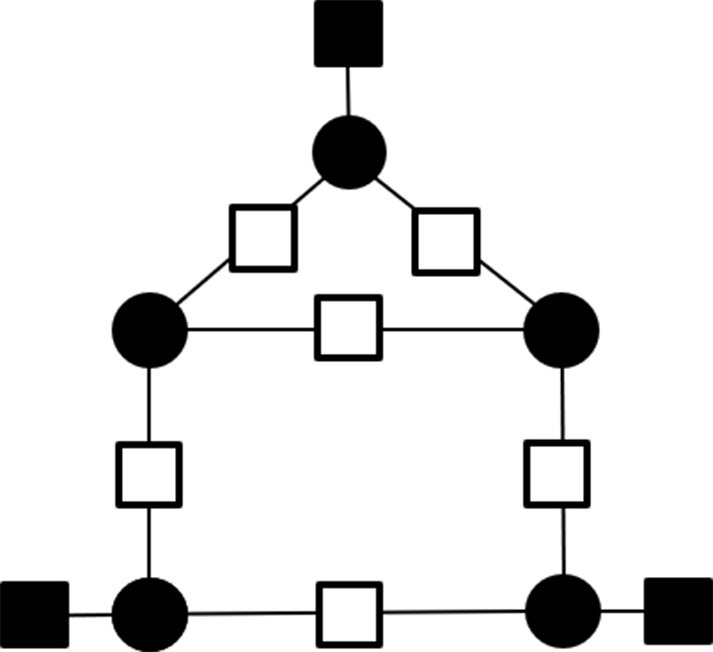}} &\sr (5,3) 
&\multirow{3}{*}{\centering \includegraphics[width=7em]{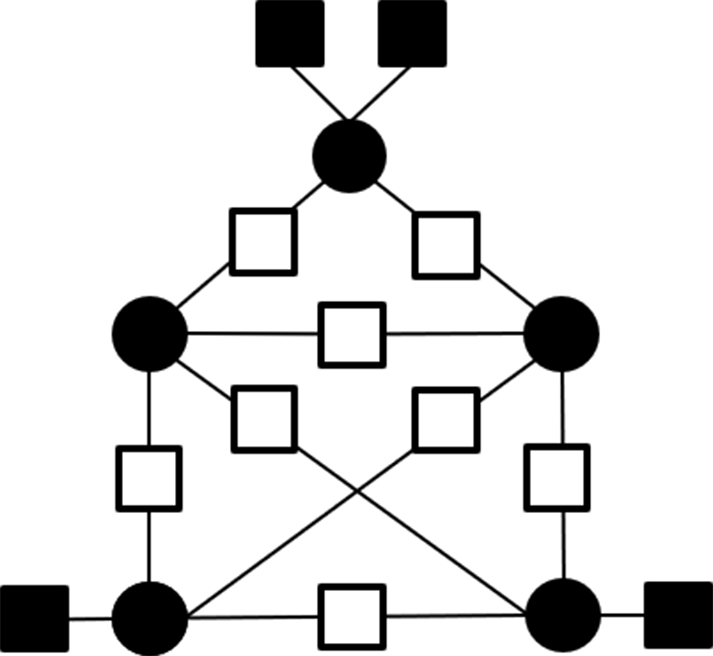}} &\sr (5,4) 
\tabularnewline
&\sr  $x^6+x^8+x^{10}$ && \sr $5x^6+5x^8+2x^{10}$ 
\tabularnewline
&\sr 4080 &&\sr 90
\tabularnewline
\hline 
\multirow{3}{*}{\centering \includegraphics[width=7em]{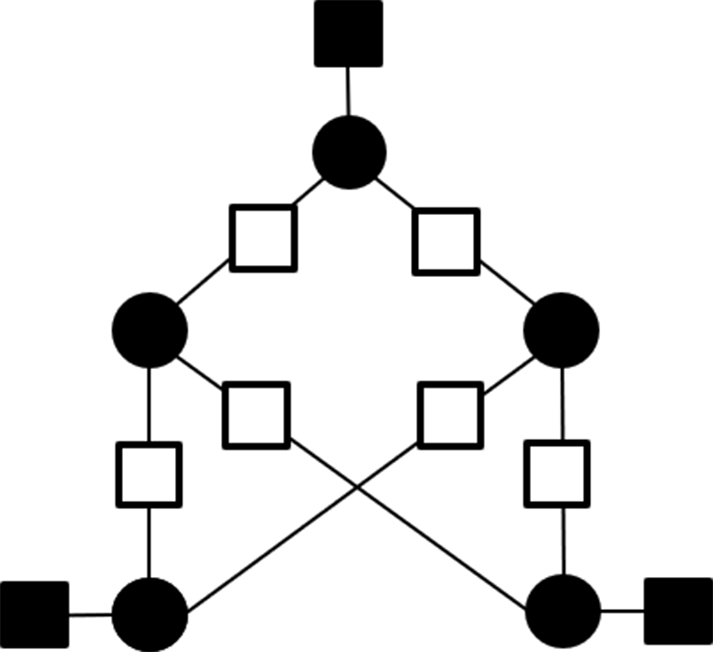}}
&\sr (5,3) &\multirow{3}{*}{\centering \includegraphics[width=7em]{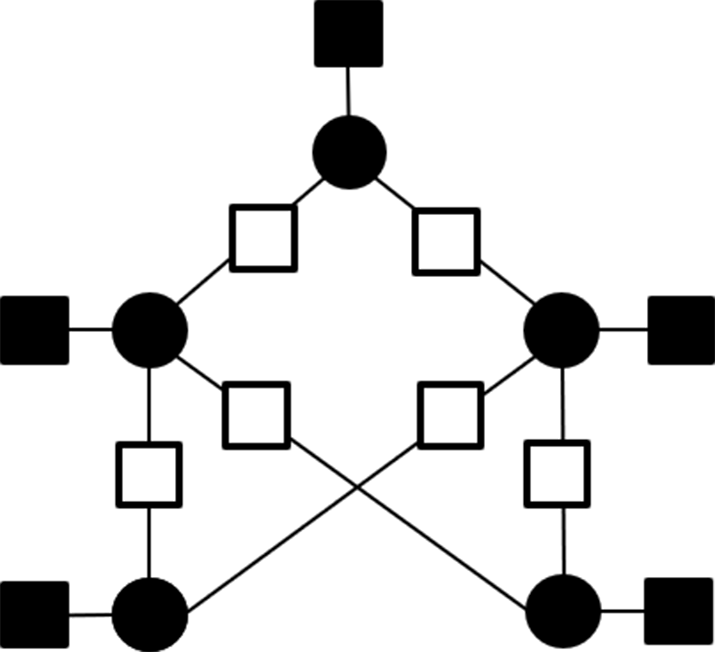}} &\sr (5,5) 
\tabularnewline
&\sr $3x^8$ &&\sr $3x^8$ 
\tabularnewline
&\sr 360 &&\sr 5184 
\tabularnewline
\hline 
\end{tabular}
\end{table*}
\footnotetext{In Table \ref{table:TSs}, the parameters-(a,b), $\textsc{CYC}(x)$, and Count are listed row-wise under the column header-Parameters for each QTS.}

\begin{figure}[ht]
	\centering
	\includegraphics[width=0.23\textwidth]{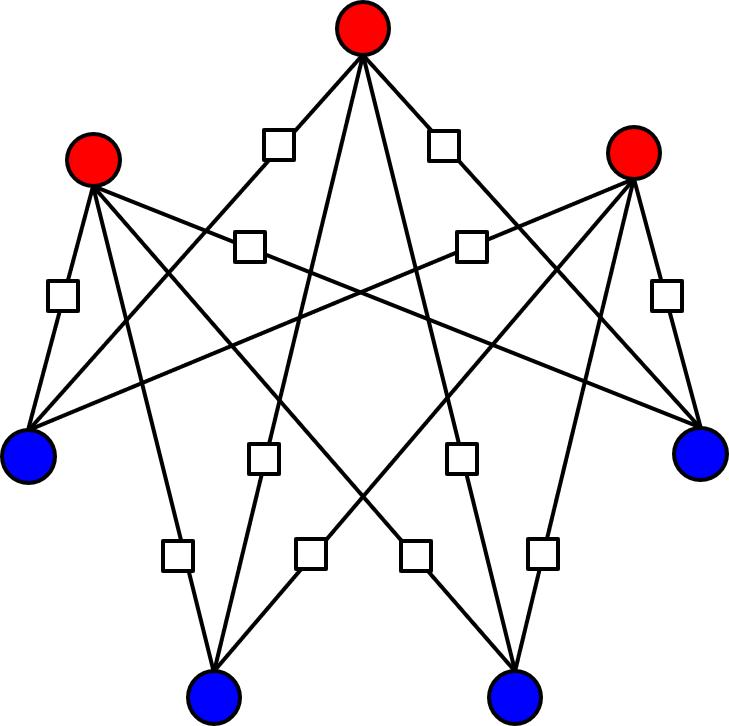}
	\caption{A stabilizer sub-graph in the $[[900, 36,10]]$ HP code \cite{roffe2020decodingQLDPC_OSD} is not symmetric.
		Note that the red and blue variable nodes have four and three check node neighbors, respectively. Thus, a BP decoder converges to the red variable nodes as its output exploiting the asymmetry in the stabilizer. }
	\label{fig:redblue_7_0_assymetricHGP}
\end{figure}

Observe that the node degree of the classical parity check Tanner graph influences the symmetric property of the stabilizers of the HP code. In Ex. \ref{ExampleHGP}, since the classical parity check code chosen has $\gamma = 3$ and $\rho = 4$ in its Tanner graph, the variable nodes of the HP code have VN degrees $3$ and $4$. For the stabilizer in Fig.~\ref{fig:redblue_7_0_assymetricHGP}, the VNs with degree $\gamma = 4$ are shown as $\color{red}{\sbullet[1.5]}$ and those with $\gamma = 3$ as $\color{blue}{\sbullet[1.5]}$. The stabilizer is not symmetric according to the Definition \ref{Def:SymmStabilizer}. Suppose the input syndrome corresponds to all the check nodes in the sub-graph in error, then an iterative BP or MSA decoder (based on the update rule) will be able to successfully converge to the red error pattern. This happens as every $\color{red}{\sbullet[1.5]}$ VN uses messages from four CNs compared to three CNs for the $\color{blue}{\sbullet[1.5]}$ VNs in the decoding process to successfully converge. Note that the above statement depends on the decoder update rule chosen. For example, the red and blue error patterns in the stabilizer are indeed failure configurations for a simple binary decoding algorithm like Gallager-B algorithm. 

This observation emphasizes the importance of the decoder in characterizing the harmfulness of QTSs as in the case of classical TSs \cite{nithinExp_ContrTCOM2020}. In the subsequent section showing the applications of TS analysis, we will use the example of different decoding schedule that ``breaks'' such symmetry of the stabilizer.

%% file: 5_SimulationResultsConclusions.tex
\section{Using the QTS to design better QLDPC codes and better decoders} 
\label{sec:SimulationResults} 
In this section, we explain practical importance of QTS analysis by providing two approaches for finite length QLDPC code and decoder design with QTS knowledge. 
\subsection{Improved Code Design}
While previous research has observed the issue of symmetric degenerate errors \cite{Rigby2019_ModifiedBP_QLDPC,Poulin_2008}, there has been no effort to fully characterize them. Identifying symmetric stabilizers, particularly of low weight in the QLDPC code is an important step in quantifying the effect of degenerate errors on iterative decoding. Removal of symmetry in low-weight stabilizers present in the Tanner graph, especially during the QLDPC code design, significantly reduces the number of instances of iterative decoder failure. For example, during the row removal step in the bicycle code \cite{mackay_quantum} we carefully modify the original bicycle code to obtain codes wherein the stabilizer is asymmetric as in Fig.~\ref{fig:degenerate_error_b}. The BP decoder will able to match to the syndrome (dark squares represent unsatisfied checks) correctly with the red error pattern, in contrast to the symmetric stabilizer in Fig.~\ref{fig:degenerate_error_a}. 
Similarly, in the case of HP codes, careful removal of TSs in the constituent classical LDPC codes helps to further optimize rate and minimum distance properties. Such code design improvements lead to performance gains especially in the error floor region for QLDPC codes. Here, we show an example of such an improved HP code design with quasi-cyclic (QC) LDPC \cite{qcCirculant_fossorier} code for the constituent classical LDPC code. 

\subsubsection{Improved HP codes without harmful TSs}
\label{subsubsec:ImprovedCodes}
One of the disadvantages of QLDPC codes is that the random code construction makes the stabilizers highly non-local, requiring arbitrary qubit-qubit inter-connectivity to perform check operations. Using QC LDPC code brings structure to the constituent codes and flexibility in improving finite length QLDPC codes along with efficient implementation of decoders. Instead of the random codes which are only optimized for girth $g=6$, we construct QC $[40,10,12]$ code with girth $8$, and make sure that small trapping sets are not present. The QC code with circulant size $Q=10$ is constructed by carefully choosing circulants to build the Tanner graph that is free of $(4,2), (4,4),$ and $(5,1)$ harmful trapping sets. For a fair comparison, we use a random code of the same size and minimum distance. The random [40,10,12] code is constructed to be cycle-4 free. The HP codes constructed from these constituent classical codes both have the same number (= 21600) of unavoidable length-8 cycles (see Lemma 3) in their respective Tanner graphs, as it only depends on the size of the component code and their variable and check node degrees. For the symmetric HP code from regular LDPC code, we can count these as $m \times \binom{\rho}{2} \times n \times \binom{\gamma}{2}$, where $m$ and $n$ correspond to the number of check nodes and variable nodes, respectively, and $d_v$ and $d_c$ are variable and check node degrees, respectively of the constituent classical code. These 8-cycles can be classified as $(4,6)$ TSs and are not harmful for iterative decoders like min-sum or BP algorithm. 

\begin{figure}[t]
	\centering
	\includegraphics[trim=12 0 20 10,clip,width=0.50\textwidth]{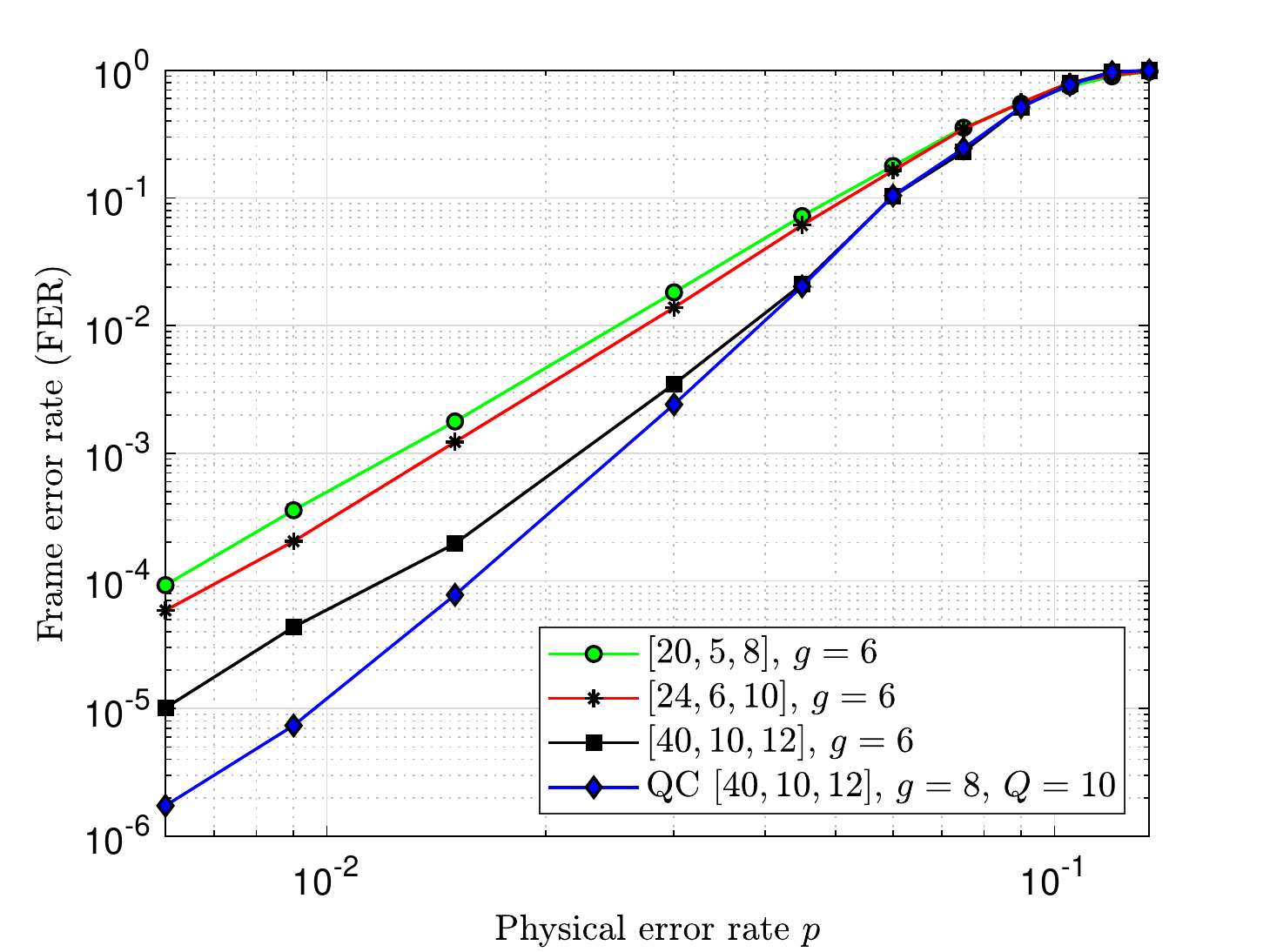}
    \caption{FER performance comparison between the symmetric HP codes constructed using random constituent codes ([20,5,8] code and [24,6,10] code in \cite{roffe2020decodingQLDPC_OSD}, and [40,10,12] code with girth $g=6$) and the symmetric HP code constructed using a trapping set aware QC [40,10,12] code. All curves are decoded using a BP decoder for $\ell_{max} = 100$ iterations with the flooding schedule. For the same code parameters, using the QC component code without harmful trapping sets shows FER improvement (close to an order of magnitude) over the random constituent code in the error floor region.}
    \label{Figure:FER_HGPC_withBP_random_vs_QC}
\end{figure}

Fig.~ \ref{Figure:FER_HGPC_withBP_random_vs_QC} shows improved decoding performance (flooding BP decoder with $\ell_{max} = 100$ iterations) in the error floor regime for the newly constructed QC HP code. Even though the curves start with a similar waterfall performance (due to same minimum distance), the randomly constructed code performs worse in the error floor regime compared to the QC code. For the same code parameters, using the QC component code improves the FER by nearly an order of magnitude at $p \approx 0.005$. This improvement can be clearly attributed to the absence of harmful TSs which were removed in the constituent QC code construction. The example of the HP-LDPC code construction with QC constituent code is chosen considering their hardware friendly nature and simple bookkeeping of trapping sets. In general, no specific code structure is required in our method, and the concept, definition and analysis of trapping sets are applicable to random codes as well.

\subsection{Novel Decoder Design} 
An alternative or complementary approach is to devise iterative decoders that do not fail for the error patterns in the QTSs identified for the QLDPC code. This approach, prevalent in classical LDPC decoders (finite alphabet iterative decoding (FAID) algorithms such as in \cite{PDDV_13_COMM}) do not ignore the topology of the TSs while devising decoder update rules. Breaking the symmetry of messages by using non-linear message update rules leads to orders of magnitude decoding error performance improvement \cite{PDDV_13_COMM}. For QLDPC iterative decoders, the typically used parallel/flooding message update schedule (in the same iteration all variable/check nodes in the Tanner graph apply in parallel the same variable/check update function, respectively) attributes to decoders' failure to symmetric degenerate errors. We devise decoder strategy that corrects these errors by taking into account the topology of the low-weight symmetric stabilizers in the code. Specifically, we show that an MSA decoder with sequential message update schedule (layered decoder as in classical literature \cite{HocevarLayered}) that uses the knowledge of location of the symmetric stabilizers in the code as well as other harmful trapping sets improves the error floor decoding performance. As an intuitive example, suppose the symmetric stabilizer of weight $6$ has support on variable nodes $v_1,\ldots,v_{6}$ with symmetric degenerate error patterns: $e_1,e_2,e_3$ and $e_4,e_5,e_{6}$. A layered decoder with the update order: starting with VN update of $v_1,v_2,v_{3}$, followed by the check node updates, and then VN update of $v_4,v_5,v_{6}$ converges to the correct error pattern without getting trapped. Since the schedule order is with respect to the variable nodes corresponding to the columns of the $H$ matrix, we refer to such schedule as column-layered. In addition to fast decoder convergence in terms of the number of iterations \cite{HocevarLayered,Sharon_SerialSchedules}, column-layered decoders break some harmful TSs in classical LDPC codes \cite{Layered_TS_ICC_2018}. In the following section \ref{subsec:ModifiedDecoders}, we compare the two schedules: flooding MSA and layered MSA decoders for the chosen QLDPC code.
\begin{figure}[t]
	\centering
	\includegraphics[trim=10 0 20 10,clip,width=0.5\textwidth]{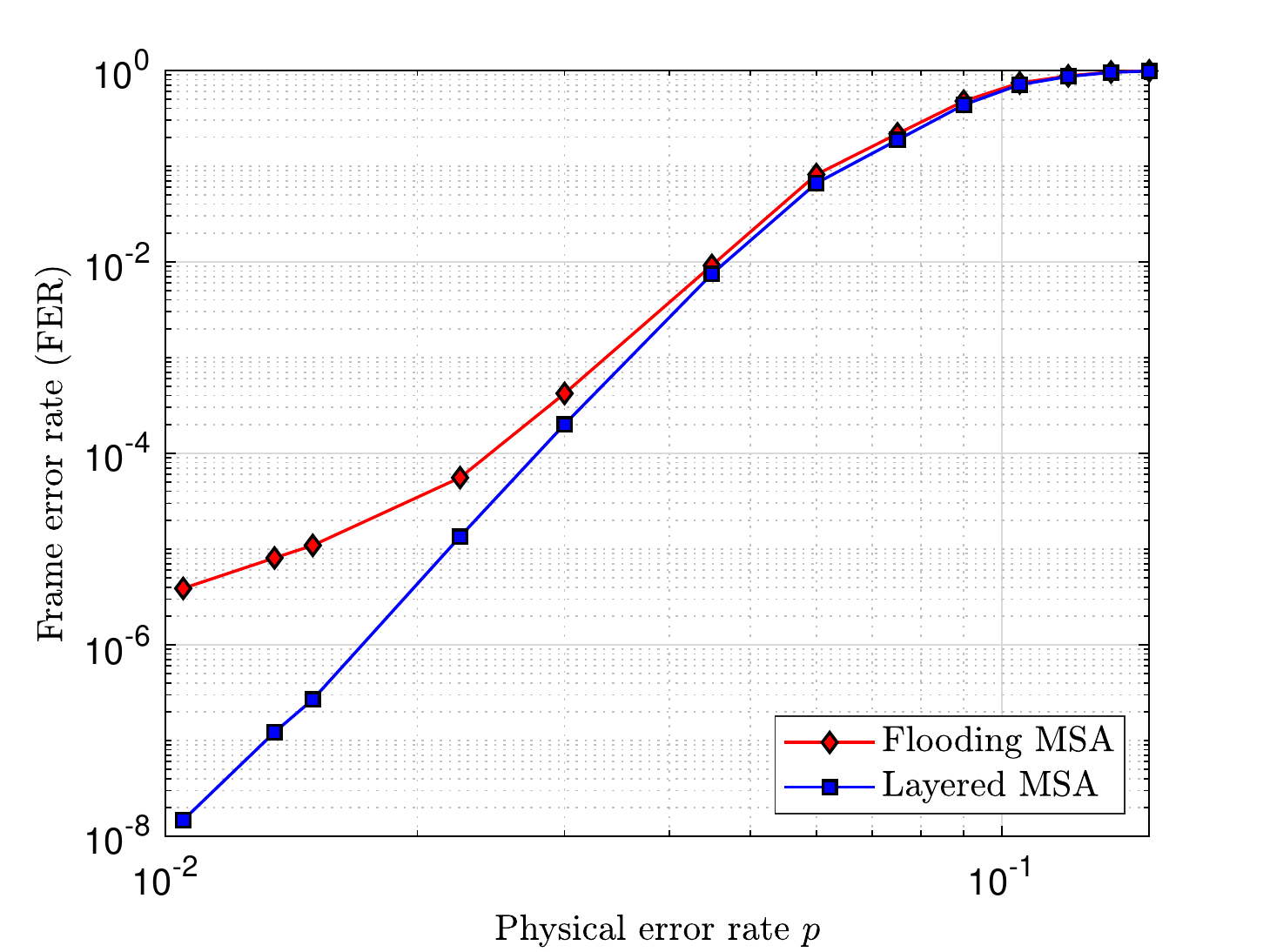}
	\caption{FER performance comparison for the $A1 [[254,28]]$ code using the min-sum algorithm (MSA) for two different schedules: flooding/parallel and layered schedule. The layered schedule is able to decode all the symmetric stabilizer TSs and numerous classical-type TSs correctly leading to two orders of magnitude improvement in the error floor regime (low physical error rates).}
	\label{Figure:GenBicycleCodeA1_hx_layered_parallel}
\end{figure}
\subsubsection{Layered Decoding to break QTSs}
\label{subsec:ModifiedDecoders}
We employ a specific layered decoding schedule to break the symmetric stabilizers in the $A1 [[254,28]]$ code in Fig.~\ref{Figure:GenBicycleCodeA1_hx_layered_parallel} using a column layered schedule. The layered schedule employed here is based on the circulant-size of the cyclic matrices $A$ and $B$. The symmetric trapping sets have a clear distinction of red and blue nodes with respect to these cyclic matrices, giving a straight forward update order: $v_1, \ldots, v_{127}$ followed by $v_{128}, \ldots, v_{254}$. The column-layered decoder (MSA with $\ell_{max} = 20$ iterations) is able to decode all the symmetric stabilizer TSs and numerous classical-type TSs correctly leading to two orders of magnitude improvement in the error floor regime (low physical error rates) compared to the flooding MSA decoder. \footnote{In the Monte-Carlo simulations presented in the paper, we collect at least 100 errors for simulation points in the low FER region ensuring that they are statistically significant.} 

\begin{figure}[t]
	\centering
	\includegraphics[trim=5 0 15 10,clip,width=0.5\textwidth]{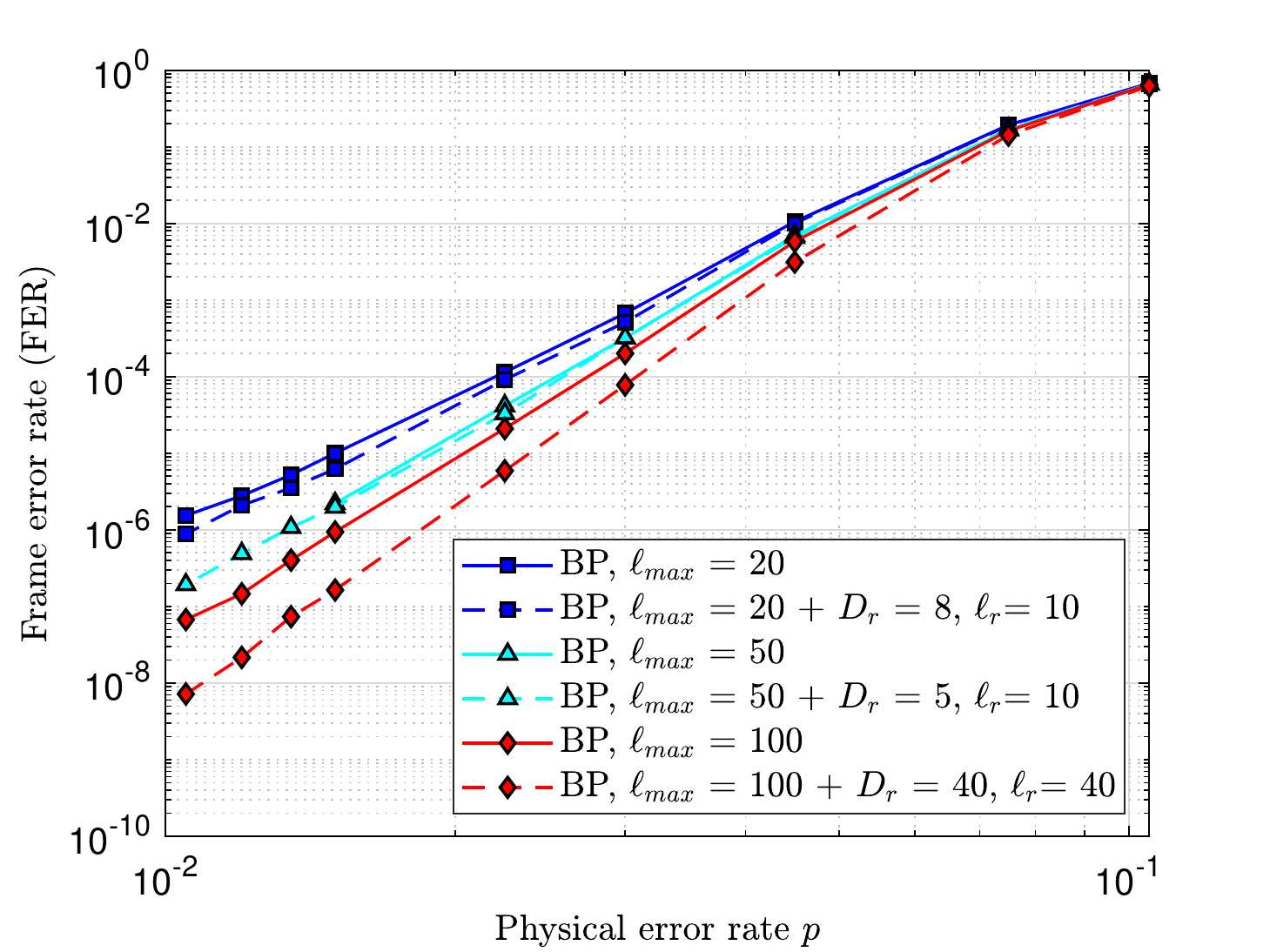}
    \caption{FER performance comparison for the $A1 [[254,28]]$ code using the belief propagation decoder with the flooding schedule with/without perturbation rounds (dashed/solid lines). When the total number of iterations allowed are only 100 iterations, the dashed-blue-square marked and dashed-cyan-triangle marked (BP with random perturbation) curves do not outperform the solid-red-diamond BP-only curve for $100$ iterations. For a significant performance improvement, post processing step with random perturbation decoders requires much more decoding rounds as shown in dashed-red-diamond curve. Also, note that the heuristic choice of noise parameter value affects these decoding improvements.}
    \label{Figure:GenBicycleCodeA1_hx_BP_randPerturb}
\end{figure}

\begin{remark}
	(Random perturbation and post-processing) \normalfont With the QTS knowledge, we can construct iterative decoders without computationally expensive post-processing steps, and thus reduce the decoding complexity and latency. Note that the two curves in Fig.~\ref{Figure:GenBicycleCodeA1_hx_layered_parallel} use same maximum number of decoding iterations ($\ell_{max} = 20$). In contrast, the heuristic random perturbation approach in \cite{Poulin_2008} requires many rounds of decoding attempts ($D_r$) with additional BP decoder iterations ($\ell_r$) each of which are initialized with perturbed channel log-likelihood ratios on variable nodes that are neighbors of the unsatisfied syndromes. If we allot the same total number of decoding iterations to compare the random perturbation method with BP, we observe that the advantage of using random perturbation is only observed with very large number of iterations. However, if the total number of iterations is set to 100 iterations, the decoding performance improvement using heuristic methods is not significant as illustrated in Fig.~\ref{Figure:GenBicycleCodeA1_hx_BP_randPerturb}. For a significant performance improvement, post processing step with random perturbation decoder requires much more decoding rounds (using $\ell_{max} + (D_r \times \ell_r) = 1700$ iterations which is $>> 100$) as shown in dashed-red-diamond curve. The heuristic choice of the randomness parameter can be further improved by using feedback as in \cite{enhanced_feedback_quant_ldpc_dec}. This is also shown to improve the convergence speed of the post-processing step, but it does not eliminate this step completely. The performance improvement shown with the chosen layered schedule is to demonstrate the usefulness of the QTS analysis. If required, post-processing techniques as well as improvements over heuristic approaches such as enhanced feedback decoding \cite{enhanced_feedback_quant_ldpc_dec} and augmented decoding \cite{Rigby2019_ModifiedBP_QLDPC} can be used as additional tools over the iterative decoder based on QTS analysis.
\end{remark}

\begin{figure}[t]
	\centering
	\includegraphics[trim=10 0 10 10,clip,width=0.5\textwidth]{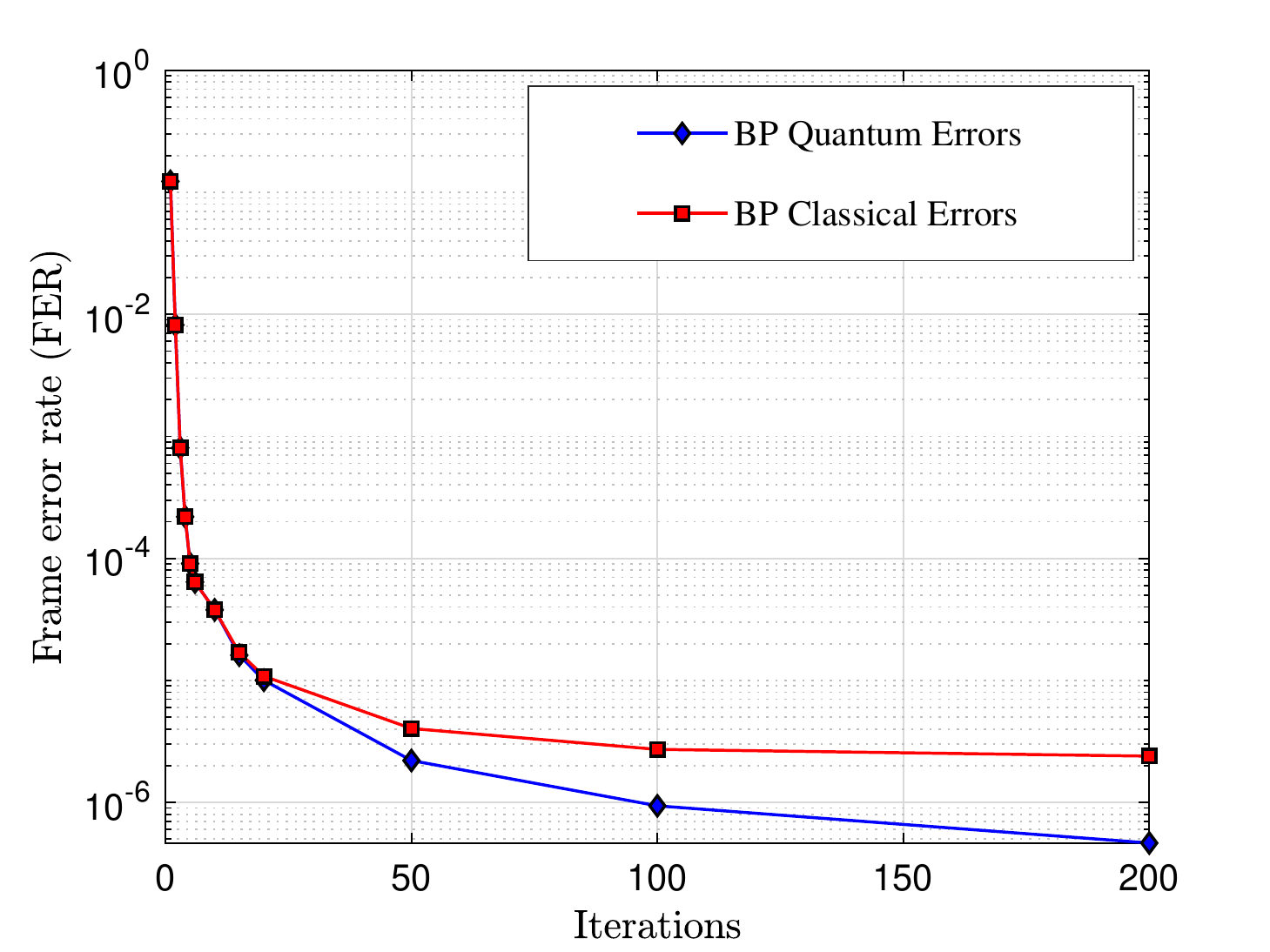}
    \caption{FER vs iterations is plotted for the physical error probability $2p/3 = 0.01$ demonstrating how error patterns are corrected by the flooding BP decoder to their degenerate patterns with more decoding iterations. The blue-diamond marked curve corresponds to the FER curve in QLDPC codes wherein degenerate error pattern matching is not considered as an error. In contrast, classical-like BP errors are plotted with the red-square marked curve with degenerate matching also classified as an error in addition to the decoding failures. Quantum decoders are observed to lower the FER with increasing iterations of BP resulting from degenerate error matching.}
    \label{Figure:GenBicycleCodeA1_FER_vsIter_01}
\end{figure}

An interesting result to note is that using more iterations in the BP algorithm results in matching with many degenerate error patterns in case of quantum LDPC codes. FER versus iterations of BP plotted in Fig.~\ref{Figure:GenBicycleCodeA1_FER_vsIter_01} shows that increasing iterations of flooding BP do not lower the FER in terms of purely classical errors (for the red curve, the degenerate error patterns that match the corresponding syndrome are considered as errors). However, such degenerate error patterns that match the syndrome without introducing a logical error are harmless for QLDPC codes, resulting in lowering  the FER with increasing iterations of BP. In our future work, we will devise improved message passing algorithms of lower complexity than BP that exploits the degeneracy of QLDPC codes systematically.

{\remark{\normalfont These alternative improved decoders are an attractive solution when the QLDPC code and their structure is fixed and modifying it is not an option (due to technology or system-level constraints in the future). Clearly, joint code and decoder design would guarantee further decoding performance improvement and a higher threshold.}}

\section{Summary and Future Work}
\label{sec:Conclusion}
In this paper, we identified and classified quantum trapping sets using their definition adapted from the classical error correction to address the syndrome decoding scenario for QLDPC codes. The knowledge of QTSs is shown to significantly improve stabilizer code/decoder designs and also decoder performance in the error floor regimes of practical finite-length QLDPC codes. Analysis of failure configurations of the QLDPC codes, which are a generalization of the surface codes, will have near-future implications in surface code designs and their decoders. 

In future work, we will analyze the finite length performances of recently proposed QLDPC codes that break the $\sqrt n$ growing minimum distance barrier \cite{panteleev2020quantumLinearMinD} based on their QTS enumeration. We will establish the parent-child relationship between the harmful sub-graphs and determine their relative harmfulness.
Understanding the effect of neighborhood of the Tanner graph with respect to the decoder used is not easy, but important to understand the actual harmful error patterns. In future work, we plan to modify the expansion-contraction method \cite{nithinExp_ContrTCOM2020} to QLDPC codes to obtain the exact set of most harmful configurations that should be avoided in the Tanner graph of QLDPC codes. Enumeration of symmetric stabilizers in QLDPC codes is also an important step towards exploiting degeneracy to the decoder's advantage. Approaches used in classical literature for structured QLDPC code constructions such as efficient low-weight codeword
search are promising in this direction. In addition, the extension of QTS definition to consider $\mathrm{X}$ and $\mathrm{Z}$ type errors together (correlated errors) and non-CSS stabilizer codes in general will set up the framework to study and explore non-binary quantum trapping sets.

%% file: 6_Appendix.tex
\appendix
\section{Iterative Decoding Algorithm}
\label{sec:AppendixA}
A syndrome-based iterative decoder $\mathcal{D}_s$ is a 6-tuple $\mathcal{D}_s =(\mathcal{M},\mathcal{Y},\zeta,\Phi,\Psi,\hat{\Phi})$,
where $\mathcal{M}$ is the message alphabets, $\mathcal{Y}$ is the same a-priori channel value chosen for all variable nodes, $\Phi$,$\Psi$ are the update functions used in variable and check nodes, and $\hat{\Phi}$ is the decision function, and $\zeta$ is the check value alphabet (for syndrome) with $\bm{\sigma}$ and $\bm{\hat{\sigma}}$ as the input and output syndromes respectively. The alphabets $\mathcal{M}$ and $\mathcal{Y}$ depend on a decoder type and quantum channel model.

Messages passed in an iterative decoder can be of floating point precision (floating point BP and MSA) or quantized to fixed number of levels for practical implementation. For a quantized decoder with $Z$ levels, the message alphabet $\mathcal{M}$ consists of $Z = 2z+ 1$ levels to which the message values are confined to. The message alphabet is defined as follows: $ \mathcal{M} = \{-B_z, \ldots ,-B_1, 0, B_1, \ldots , B_z\},$ where $B_i \in \mathbb{Z}+$ (positive integers) and $B_i > B_j$ for any $i > j$. The sign of a message $m \in \mathcal{M}$ can be interpreted as the error estimate of the variable node for which $m$ is being passed to or from (positive for zero and negative for one), and the magnitude as a measure of how reliable the error estimate is. For BSC, the initial channel value for variable node $v_i$ is set as $y_i = +Y$ mapping $0 \rightarrow Y$ according to the assumption of zero error pattern. The variable node message from $v_i$ is initialized to $\Phi(y_i,\mathbf{0})$, and in each iteration updated according to the rules $\Phi$ and $\Psi$.

The messages passed over the edges of the Tanner graph (say, at $\ell$-th iteration-iteration will be indicated as superscript when required) are denoted as follows:
$\mu_{c_i \to v_j}$ and $\nu_{v_j \to c_i}$ denote a message from check node $c_i$ to variable node $v_j$ and vice-versa respectively. 

Check node message is updated as $\mu^{(\ell)}_{c_i \to v_j}= \Psi(\mathbf{n}^{(\ell-1)}, \sigma_i)$, where $\mathbf{n}=\nu_{\mathcal{N}(c_i) \backslash v_j \to c_i}$ denote all incoming variable node messages to the check node $c_i$ except from the variable node $v_j$. Note that $\Psi$ is a symmetric function, i.e., any permutation of the function variables leaves the function unchanged. Variable node message is updated as $\nu^{(\ell)}_{v_j \to c_i} = \Phi(y_j,\mathbf{m}^{(\ell)})$, where $\mathbf{m}=\mu_{\mathcal{N}(v_j) \backslash c_i \to v_j}$ denote all incoming check node messages to the variable node $v_j$ except the message from the check node $c_i$.

The decision function on $v_j$ is computed using all messages incoming to $v_j$ denoted by $\mathbf{l} = \mu_{\mathcal{N}(v_j) \to v_j}$. The decision function $\lambda_j^{(\ell)} = \hat{\Phi} (\mathbf{l}^{(\ell)}, y_j)$ decides the error bit $\hat{E}_j$ based on the sign using an indicator function as $\hat{e}_{j}^{(\ell)} = \mathbbm{1}_{\lambda_{j}^{(\ell)}<0}$. Output syndrome value for $i^{th}$ check node in the $\ell$-th iteration $\hat{\sigma}_i^{(\ell)}= \sum_{k \in \mathcal{N}(c_i)} \hat{e}_k$ modulo-2. A check node $c_i$ is matched only if $\hat{\sigma}_i=\sigma_i$. If all syndromes are matched, we say that iterative decoder $\mathcal{D}_s$ successfully decoded to output the error pattern $\hat{\bf{e}}$. 

The order of message passing in the Tanner graph is generally referred to as the updating schedule. Message passing follows a parallel/flooding schedule where $\Psi$ at all CNs are updated simultaneously followed by updating $\Phi$ at all VNs simultaneously. In contrast, a row (column) layered schedule performs sequential update of messages in an order. An iteration of row (column) layered decoder proceeds by computing a check node (variable node) update function in the sequence followed by computing neighboring variable node (check node) function till all check nodes (variable nodes) are updated. Decision function $\hat{\Phi}$ computed at each layer accelerates the decoder convergence significantly.